\documentclass[useAMS,usenatbib]{mn2e}
\usepackage [english]{babel}
\usepackage{latexsym}
\usepackage{epsfig}
\usepackage[T1]{fontenc}
\usepackage{graphicx}
\usepackage{fancyhdr}
\usepackage{indentfirst}
\usepackage{subfigure}
\usepackage{amsmath}
\usepackage{amsfonts}
\usepackage{hyperref}
\usepackage{placeins}

\title[Classification of astronomical transients]{An analysis of feature relevance in the classification of astronomical transients with machine learning methods}
\author[D'Isanto et al. 2015]{A. D'Isanto$^{1, 2}$\thanks{E-mail: antonio.disanto@h-its.org}, S. Cavuoti$^{3}$, M. Brescia$^{3}$, C. Donalek$^{4}$, G. Longo$^{1}$, G. Riccio$^{3}$, \and S. G. Djorgovski$^{4, 5}$.\\
 $^{1}$Department of Physical Sciences, University of Napoli Federico II, via Cinthia 9, 80126 Napoli, ITALY\\
 $^{2}$Heidelberg Institute for Theoretical Studies (HITS), Schloss-Wolfsbrunnenweg 35, 69118 Heidelberg, GERMANY\\
 $^{3}$INAF - Astronomical Observatory of Capodimonte, via Moiariello 16, 80131 Napoli, ITALY\\
 $^{4}$Center for Data Driven Discovery, California Institute of Technology, 1200 E. California Blvd., 91125 Pasadena, USA\\
 $^{5}$Department of Astronomy, California Institute of Technology, 1216 East California bvd., Pasadena, CA 91125, USA\\
 }

\date{Accepted . Received ; in original form }

\pagerange{\pageref{firstpage}--\pageref{lastpage}} \pubyear{2015}
\begin{document}
\label{firstpage}
\maketitle
\begin{abstract}
The exploitation of present and future synoptic (multi-band and multi-epoch) surveys requires an extensive use of
automatic methods for data processing and data interpretation.  In this work, using data extracted from the
Catalina Real Time Transient Survey (CRTS), we investigate the classification performance of some well tested methods:
Random Forest, MLPQNA (Multi Layer Perceptron with Quasi Newton Algorithm) and K-Nearest Neighbors, paying special attention to
the feature selection phase.
In order to do so, several classification experiments were performed. Namely:
identification of cataclysmic variables, separation between galactic and extra-galactic objects and identification of supernovae.
\end{abstract}

\section{Introduction}

The advent of a new generation of multi-epoch and multi-band (synoptic) surveys has opened a new era in astronomy allowing to study with unprecedented accuracy the physical properties of variable sources.
The potential of these new digital surveys, both in terms of new discoveries as well as of a better understanding of already known phenomena, is huge.
For instance, the Catalina Real-Time Transient Survey (CRTS, \citealt{drake}) in less than $8$ years of operation, enabled the discovery of $\sim2400$ \textit{SN}, $\sim1200$ \textit{CV}, $\sim2800$ \textit{AGN}, as well as to identify brand new phenomena such as binary black holes \citep{graham2015} and peculiar types of supernovae \citep{drake2010}.
A discovery trend which is expected to continue and even increase when new observing facilities such as the Large Synoptic Telecope (LSST, \citealt{closson2015}), and the Square Kilometer array (SKA, \citealt{yahya2015}) become operational.

With these new instruments, however, both size of the data and event discovery rates are expected to increase, from the current $\sim10-10^{2}$ events per night, up to $\sim10^{5}-10^{7}$.
Only a small fraction of these events will be targeted by dedicated follow-up's and therefore it will become crucial to disentangle potentially interesting events from lesser ones.
With data volumes already in the terabyte and petabyte domain, the discrimination of time-critical information has already exceeded the capabilities of human operators and also crowds of
\textit{citizen scientists} cannot match the task.
A viable approach is therefore to automatize each step of the data acquisition, processing and understanding tasks. In this work, with ''data understanding'' we mean the identification of transients and their classification into broad classes, such as periodic vs non periodic, supernovae, Cataclysmic Variables (\textit{CV}) stars, etc.

Many efforts have been made to apply a variety of machine learning (ML) methods to classification problems \citep{dubuisson2014, goldstein2015, rebbapragada2014, wright}.

Real time analysis can be performed using different methods, among which we shall just recall those based on Random Forest (RF; \citealt{breiman2001}) and on Hierarchical Classification \citep{kitty}.

Off-line classification, being less critical in terms of computing time, can be performed with many different types of classifiers. It is common practice to distinguish between supervised and unsupervised
methods, depending on whether a previously classified sample is or is not used for the training phase. In the \textit{supervised} category we have, for instance, Bayesian Network \citep{castillo},
Support Vector Machines (SVM, \citealt{chang2011}), K-nearest neighbors (KNN, \citealt{hastie}), Random Forest (\citealt{breiman2001}), and Neural Networks \citep{pitts}.
While in the unsupervised family we mention Gaussian Mixture Modeling (GMM, \citealt{McLachlan}), and Self-Organizing Maps (SOM, \citealt{kohonen2007}).

In this work we shall focus on off-line classification, making use of three different machine learning methods, namely: the Multi Layer Perceptron with Quasi-Newton Algorithm (MLPQNA, \citealt{brescia2}), the Random Forest (RF, \citealt{breiman2001}) and the K-Nearest Neighbors (KNN, \citealt{hastie}). Most of the presented work was performed in the framework of the Data Mining \& Exploration Web Application REsource (DAMEWARE, \citealt{dame}) infrastructure and the PhotoRaptor public tool \citep{cavuoti2015}.

\begin{figure}
\begin{center}
\includegraphics[width=8cm]{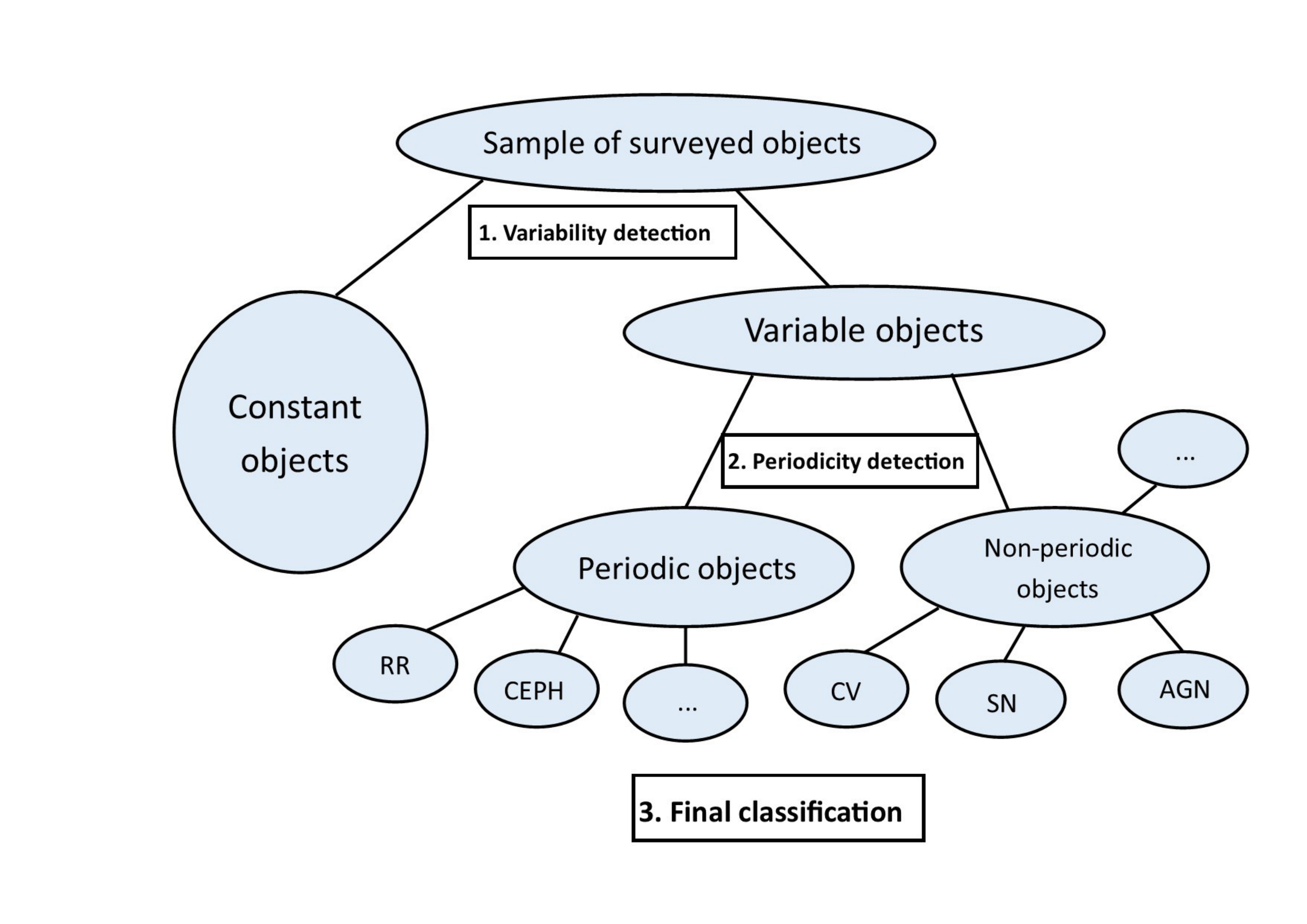}
\caption[Modified Dubath classification scheme.]{An adapted version of the scheme presented in \cite{dubath1} for a general classification of variable objects.} \label{class_ext}
\end{center}
\end{figure}

The paper is structured as it follows: in Section \ref{SEC:thedata} we present the data and introduce the features extracted for the analysis.
In Section \ref{SEC:themethods} we briefly describe the machine learning methods used for the experiments detailed in Section \ref{SEC:experiments}. Results are discussed in \ref{SEC:discussion}.

%%%%%%%%%%%%%%%%%%%%%%%%%%%%%%%%%%%
\section{The Data}\label{SEC:thedata}

In what follows we shall divide objects according to a simplified version (see Fig. \ref{class_ext}) of the semantic tree described in \cite{eyer}.
From this scheme it emerges quite naturally, the need to split the classification task in at least three steps (e.g. \citealt{dubath1}).
In the first step, variable objects (the transients) are disentangled from normal, non variable stars.
In the second step, periodic objects are separated from non-periodic objects and, finally, in the third and last step, one can proceed to the final classification of the objects.

In this work we make use of $1,619$ light curves extracted from the Catalina Real-Time Transient Survey (CRTS, \citealt{drake}) public archive. CRTS is a synoptic astronomical survey that repeatedly covers thirty three thousand square degrees of the sky with the main goal of discovering rare and interesting transient phenomena. The survey utilizes data taken in only one band (V) by the three dedicated telescopes of the highly successful Catalina Sky Survey (CSS) NEO project and detects and openly publishes all transients within minutes of observation so that all astronomers may follow ongoing events.

\noindent The sample used in the present work consists of the light curves of objects whose nature was confirmed with spectroscopic or photometric follow-up's, and it is composed by:

\begin{itemize}
 \item{Cataclysmic Variables - \textit{CV} ($461$ objects);}
 \item{Supernovae - \textit{SN} ($536$ objects);}
 \item{Blazar - \textit{Bl} ($124$ objects);}
 \item{Active Galactic Nuclei - \textit{AGN} ($140$ objects);}
 \item{Flare Stars - \textit{Fl} ($66$ objects);}
 \item{RR Lyrae - \textit{RRL} ($292$ objects).}
\end{itemize}

\subsection{Photometric features}
\label{features}

The ability to recognize and quantify the differences between light curves with ML methods, requires many instances of light curves for each class of interest.
As extensively discussed (cf. \citealt{donalek2,bloom,graham,wright}), in analysing astronomical time series, it is crucial to extract from the light curves a proper set of features.
Since light curves are usually unevenly sampled, and not all instances of a certain class are observed with the same number of epochs and S/N ratio, the use of the light curves themselves
for classification purposes is therefore challenging, both conceptually and computationally.
Therefore, the data need to be homogenized by transforming each light curve into a vector of real-number features generated using statistical and/or model-specific fitting procedures.

In this work we used the Caltech Time Series Characterization Service (CTSCS), a publicly offered web service \citep{graham2012}, to derive from a given light curve a rather complete set
of features capable to characterize both periodic \citep{richards,debosscher} and non periodic behaviors.

Among the many possible features provided by the service, we used those listed below.

\begin{itemize}
\item\underline{Amplitude} ($ampl$): the arithmetic average between the maximum and minimum magnitude;
\begin{equation}
ampl = \frac{mag_{max} - mag_{min}}{2}
\end{equation}

\item \underline{Beyond1std} ($b1std$): the fraction of photometric points ($\leq 1$)  above or under a certain standard deviation from the weighted average (by photometric errors);
\begin{equation}
b1std = P(|mag - \overline{mag}| > \sigma)
\end{equation}

\item \underline{Flux Percentage Ratio} ($fpr$): the percentile is the value of a variable under which there is a certain percentage of light curve data points. The flux percentile $F_{n,m}$ was defined as the difference between the flux values at percentiles $n$ and $m$.
The following flux percentile ratios have been used:

\begin{description}
\item{$fpr20 = F_{40,60}/F_{5,95}$}
\item{$fpr35 = F_{32.5,67.5}/F_{5,95}$}
\item{$fpr50 = F_{25,75}/F_{5,95}$}
\item{$fpr65 = F_{17.5,82.5}/F_{5,95}$}
\item{$fpr80 = F_{10,90}/F_{5,95}$}
\end{description}

\item \underline{Lomb-Scargle Periodogram} ($ls$): the period obtained by the peak frequency of the Lomb-Scargle periodogram  \citep{scargle};

\item \underline{Linear Trend} ($lt$): the slope of the light curve in the linear fit, that is to say the $a$ parameter in the following linear relation:
\begin{equation}
mag = a*t + b
\end{equation}

\begin{equation}
lt = a
\end{equation}

\item \underline{Median Absolute Deviation} ($mad$): the median of the deviation of fluxes from the median flux;
\begin{equation}
mad = median_{i}(|x_{i} - median_{j}(x_{j})|)
\end{equation}

\item \underline{Median Buffer Range Percentage} ($mbrp$): the fraction of data points which are within $10\%$ of the median flux;
\begin{equation}
mbrp = P(|x_{i} - median_{j}(x_{j})| < 0.1*median_{j}(x_{j}))
\end{equation}

\item \underline{Magnitude Ratio} ($mr$): an index used to estimate if the object spends most of the time above or below the median of magnitudes;
\begin{equation}
mr = P(mag > median(mag))
\end{equation}

\item \underline{Maximum Slope} ($ms$): the maximum difference obtained measuring magnitudes at successive epochs;
\begin{equation}
ms = max(|\frac{(mag_{i+1} - mag_{i})}{(t_{i+1} - t_{i})}|) = \frac{\Delta mag}{\Delta t}
\end{equation}
	
\item \underline{Percent Amplitude} ($pa$): the maximum percentage difference between maximum or minimum flux and the median;
\begin{equation}
pa = max(|x_{max} - median(x)|, |x_{min} - median(x)|)
\end{equation}

\item \underline{Percent Difference Flux Percentile} ($pdfp$): the difference between the second and the $98th$ percentile flux, converted in magnitudes. It is calculated by the ratio $F_{5,95}$ on median flux;
\begin{equation}
pdfp = \frac{(mag_{95} - mag_{5})}{median(mag)}
\end{equation}

\item \underline{Pair Slope Trend} ($pst$): the percentage of the last $30$ couples of consecutive measures of fluxes that show a positive slope;
\begin{equation}
pst = P(x_{i+1} - x_{i} > 0, i = n-30,...,n)
\end{equation}
	
\item \underline{R Cor Bor} ($rcb$): the fraction of magnitudes that is below 1.5 magnitudes with respect to the median;
\begin{equation}
rcb = P(mag > (median(mag) + 1.5))
\end{equation}

\item \underline{Small Kurtosis} ($sk$): the kurtosis represents the departure of a distribution from normality and it is given by the ratio between the 4th order momentum and the square of the variance.
For small kurtosis it is intended the reliable kurtosis on a small number of epochs;
\begin{equation}
sk = \frac{\mu_{4}}{\sigma^{2}}
\end{equation}

\item \underline{Skew} ($skew$): the skewness is an index of the asymmetry of a distribution. It is given by the ratio between the 3rd order momentum and the variance to the third power;
\begin{equation}
skew = \frac{\mu_{3}}{\sigma^{3}}
\end{equation}
	
\item \underline{Standard deviation} ($std$): the standard deviation of the fluxes.
\end{itemize}

%%%%%%%%%%%%%%%%%%%%%%%%%%%%%%%%
\section{The methods}\label{SEC:themethods}

As it was said before, this work aims to classify transients using a machine learning approach based on the use of various methods: MLPQNA, RF and KNN.

MLPQNA stands for the classical Multi-Layer Perceptron model implemented with a Quasi Newton Approximation (QNA) as learning rule \citep{byrd1994}.
This model has already been used to deal with astrophysical problems and it is extensively described elsewhere \citep{brescia2,cavuoti2014}.

RF stands instead for Random Forest, a widely known ensemble method \citep{breiman2001}, which uses a random subset of data features to build an ensemble of decision trees.
Our implementation makes use of the public library scikit-learn \citep{pedregosa}.
This method has been chosen mainly because it provides for each input feature a score of importance (rank) measured in terms of its contribution percentage to the classification results.

KNN is the well known k-Nearest Neighbors method \citep{hastie}, widely used both for classification and regression.
In the case of classification, it tries to classify an object by a majority vote of its neighbors, and the object is then assigned to the most common class among its k nearest neighbors.

The analysis of the results of the experiments is based on the so-called confusion matrix \citep{provost1998}, a widely used classification performance visualization matrix,
where columns represent the instances in a predicted class, and rows give the expected instances in the known classes.
In a confusion matrix defined as in Tab.~\ref{conf_matrix} the quantities are:
$TP$: true positive,
$TN$: true negative,
$FP$: false positive,
$FN$: false negative.

By combining such terms, it is then possible to derive the following statistical parameters (in brackets the label that will be used in the tables):

\begin{itemize}
\item \underline{overall Efficiency} (\textit{Eff}): the ratio between the number of correctly classified objects and the total number of objects in the data set;
\begin{equation}
Eff = \frac{TP + TN}{TP + FP + FN + TN}
\end{equation}

\item \underline{class Purity} (\textit{Pur1} and \textit{Pur2}): the ratio between the number of correctly classified objects of a class and the number of objects classified in that class, also known as efficiency of a class;
\begin{equation}
Pur1 = \frac{TP}{TP + FP}
\end{equation}
\begin{equation}
Pur2 = \frac{TN}{FN + TN}
\end{equation}

\item \underline{class Completeness} (\textit{Comp1} and \textit{Comp2}): the ratio between the number of correctly classified objects in that class and the total number of objects of that class in the data set;
\begin{equation}
Comp1 = \frac{TP}{TP + FN}
\end{equation}
\begin{equation}
Comp2 = \frac{TN}{FP + TN}
\end{equation}

\item \underline{class Contamination}: it is the dual of the purity. Namely it is the ratio between the number of misclassified object in a class and the number of objects classified in that class.
Since easily derivable from the purity percentages, it is not explicitly listed in the results;

\item \underline{Matthews Correlation Coefficient} (\textit{MCC}): it is an index used as a quality measure for a two-class classification. It takes into account values derived from the confusion matrix, and can be used also if the classes are very unbalanced. It can be regarded as a correlation coefficient between the observed and predicted binary classification, returning a value between -1 and 1. Where -1 indicates total disagreement between prediction and observation, 0 indicates random prediction, and 1 stands for a perfect prediction \citep{Matthews}.

\begin{equation}\footnotesize
MCC = \frac{TP \times TN - FP \times FN}{\sqrt{(TP + FP)(TP + FN)(TN + FP)(TN + FN)}}
\end{equation}

\end{itemize}

These parameters can be used to describe completely the distribution of the blind test patterns after training.

Moreover, in order to compare the three classifiers used, we also derived the Receiver Operating Characteristic or ROC curve plots for the most significant experiments.
A ROC curve is a graphical diagram showing the classification performance trend by plotting the true positive rate against the false positive rate as the classification threshold is varied \citep{hanley}.
The overall effectiveness of the algorithm is measured by the area under the ROC curve, where an area of $1$ represents a perfect classification, while an area of .5 indicates a useless result.

\begin{table}\scriptsize
\centering
\begin{tabular}{| c | c | c | c |}
 & &\textbf{OUTPUT}& \\
 &-&class 1&class 2\\
\hline
\hline
\textbf{TARGET}&class 1&$TP$&$FN$\\
 &class 2&$FP$&$TN$\\
\hline
\end{tabular}
\caption{Structure of the confusion matrix for a two classes experiment. The interpretation of the symbols is self explanatory. For instance, $TP$ denotes the number of objects belonging to the class 1 who are correctly classified.}
\label{conf_matrix}
\end{table}

\section{Classification experiments}\label{SEC:experiments}

We performed the following classification experiments:

\begin{itemize}
\item multi-class (\textit{six-class}), in which the whole catalog, including all the six classes, was separately considered, in order to investigate the capability to correctly disentangle at once all the given categories of variable objects;

\item Cataclismic Variables \textit{(CV) vs ALL}, where the category \textit{ALL} includes \textit{AGN, SN, Fl, Bl} types. Here the \textit{RRL} type was not considered;

\item Extra-Galactic (\textit{AGN} and \textit{Bl} types) vs Galactic (\textit{CV}, \textit{SN} and \textit{Fl} types), to search for an improvement with respect to the previous separation.
The inclusion of \textit{SN} type in the Galactic class is motivated by the fact that, even though mainly observed in external galaxies, they are stars and therefore represent a completely different
category with respect to active galactic nuclei;

\item \textit{SN vs ALL}, where \textit{ALL} includes \textit{AGN, Bl, CV, Fl} and \textit{RRL} types.
\end{itemize}

For each classification experiment we adopted the same strategy.
First of all, we run a RF experiment using all $20$ features described in Sec.~\ref{features}, in order to obtain a feature importance ranking (i.e. the relevance of each feature to the classification expressed in terms of information entropy).
The results of the RF experiment allowed us to select different groups of features (ordered by ranking), to be used for a second set of binary classification experiments performed with MLPQNA, RF and KNN.
Finally, using the best set of features, we performed an heuristic optimization of the MLPQNA parameters (i.e. complexity of the network topology as well as the Quasi-Newton learning decay factor), aimed at improving the classification results.

We then froze the topology of the MLPQNA using $1$ hidden layer, while for the RF we chose a $10,000$ trees configuration, and finally for the KNN we chose $k = 5$.
Moreover, we always applied a $10$-fold cross validation \citep{geisser1975}, in order to obtain statistically more robust results (i.e. to avoid any potential occurrence of overfitting in the training phase). In terms of performance evaluation, it is important to underline that we were mostly interested to the classification purity percentages. Therefore these indicators have been primarily evaluated to assign the best results.

\subsection{Multi-class}

\begin{figure*}
\begin{center}
\includegraphics[width=15cm]{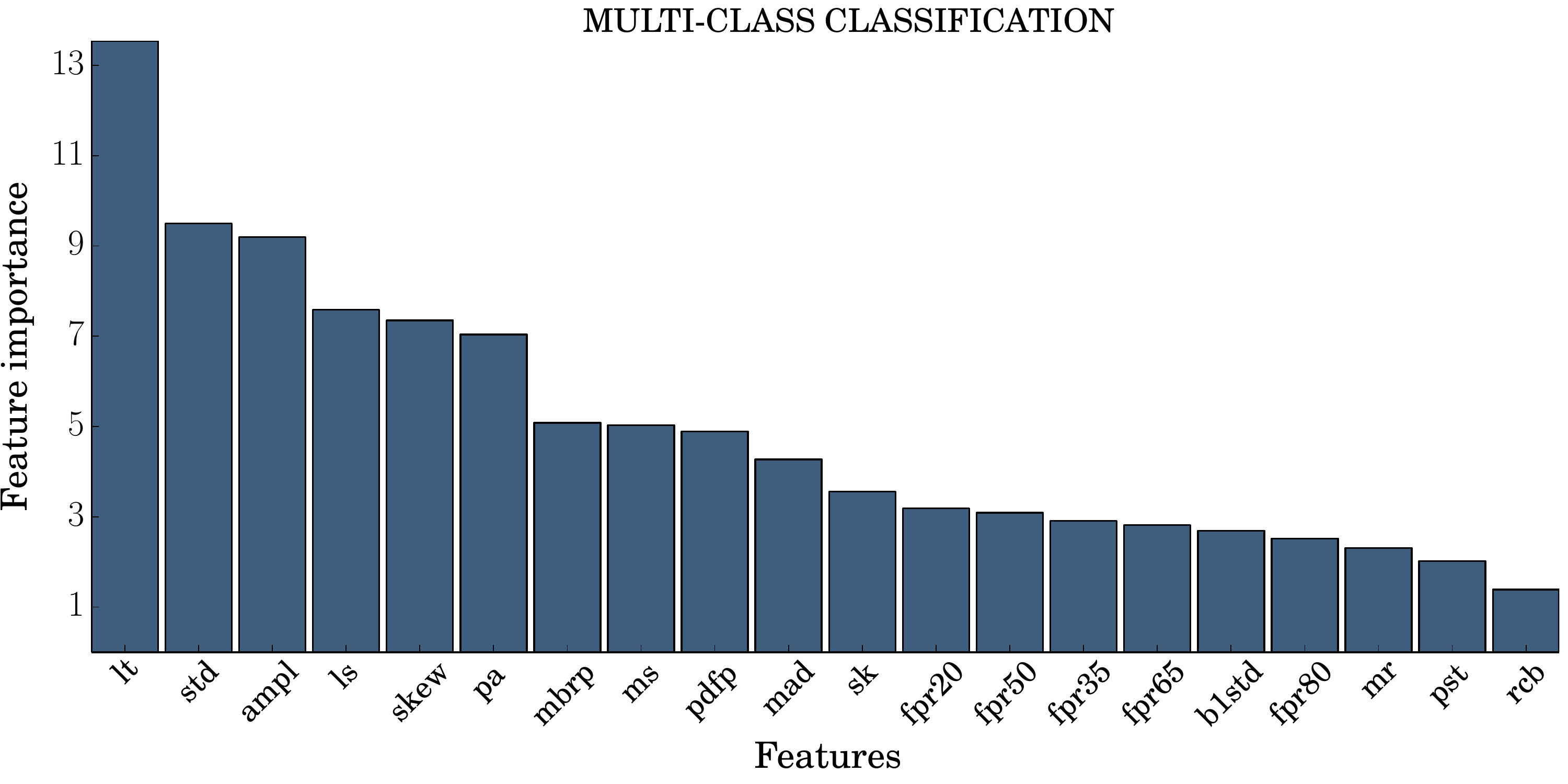}
\caption[Feature importance list obtained with RF for the \textit{six-class} experiment.]{Feature importance list obtained by the RF in the case of the \textit{six-class} experiment, with the importance percentage for each feature.} \label{list_six}
\end{center}
\end{figure*}

\begin{figure*}
     \begin{center}

        \subfigure[\textit{CV}]{
            \label{6class:cv}
            \includegraphics[width=0.48\textwidth]{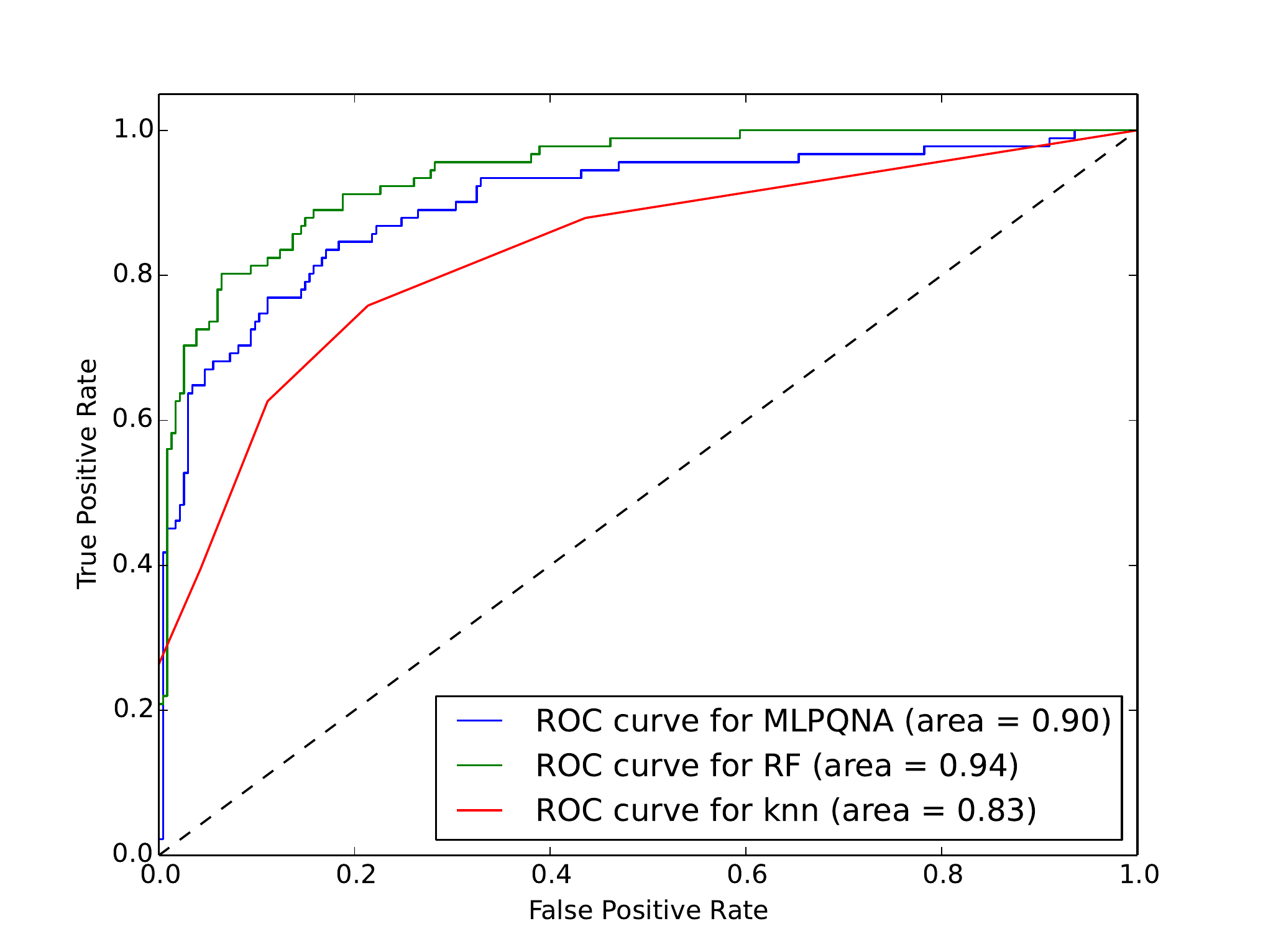}

        }
        \subfigure[\textit{SN}]{
           \label{6class:sn}
           \includegraphics[width=0.48\textwidth]{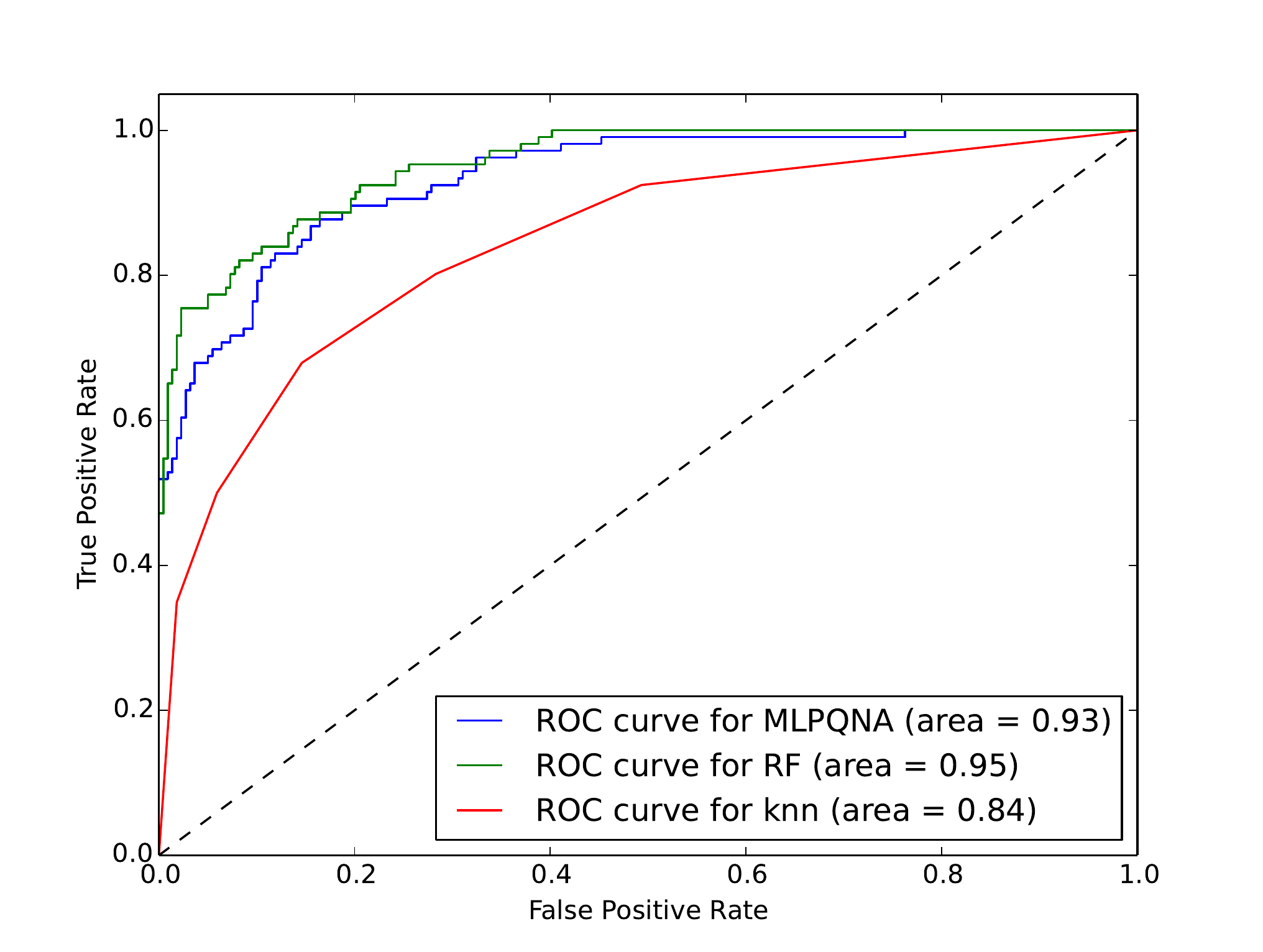}

        }\\
        \subfigure[\textit{Bl}]{
            \label{6class:bl}
            \includegraphics[width=0.48\textwidth]{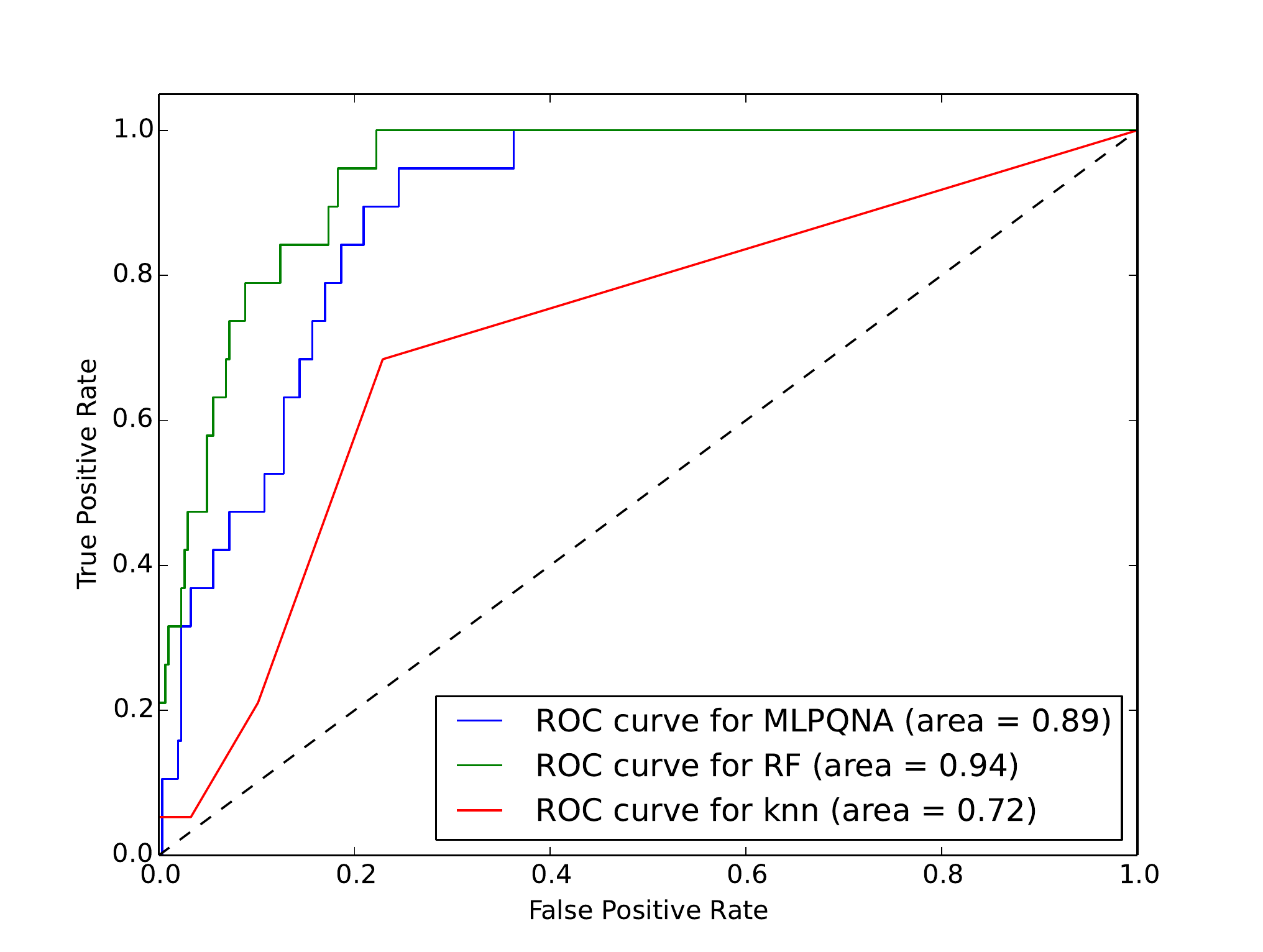}
        }
        \subfigure[\textit{AGN}]{
            \label{6class:agn}
            \includegraphics[width=0.48\textwidth]{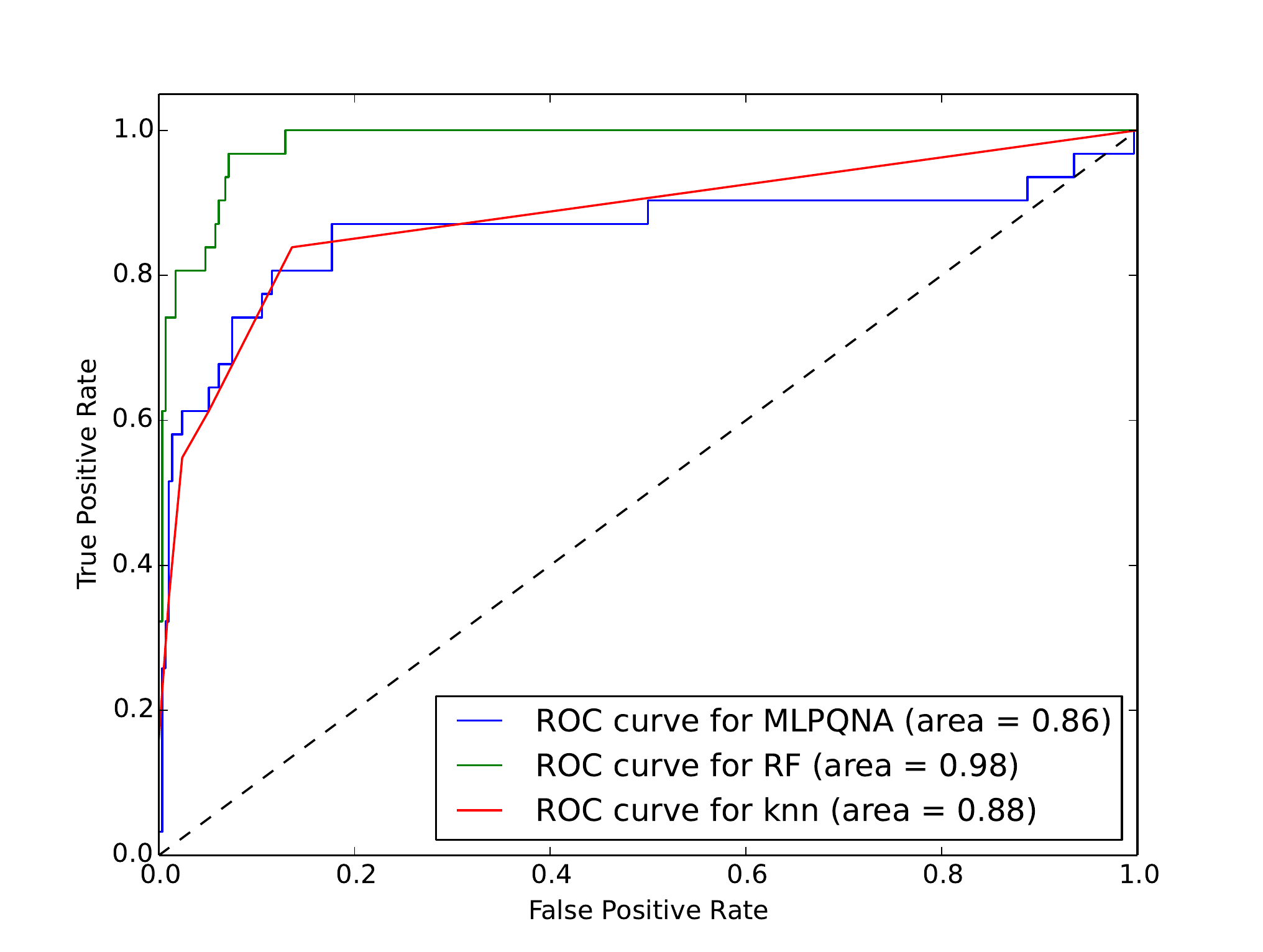}

        }\\
        \subfigure[\textit{Fl}]{
           \label{6class:fl}
           \includegraphics[width=0.48\textwidth]{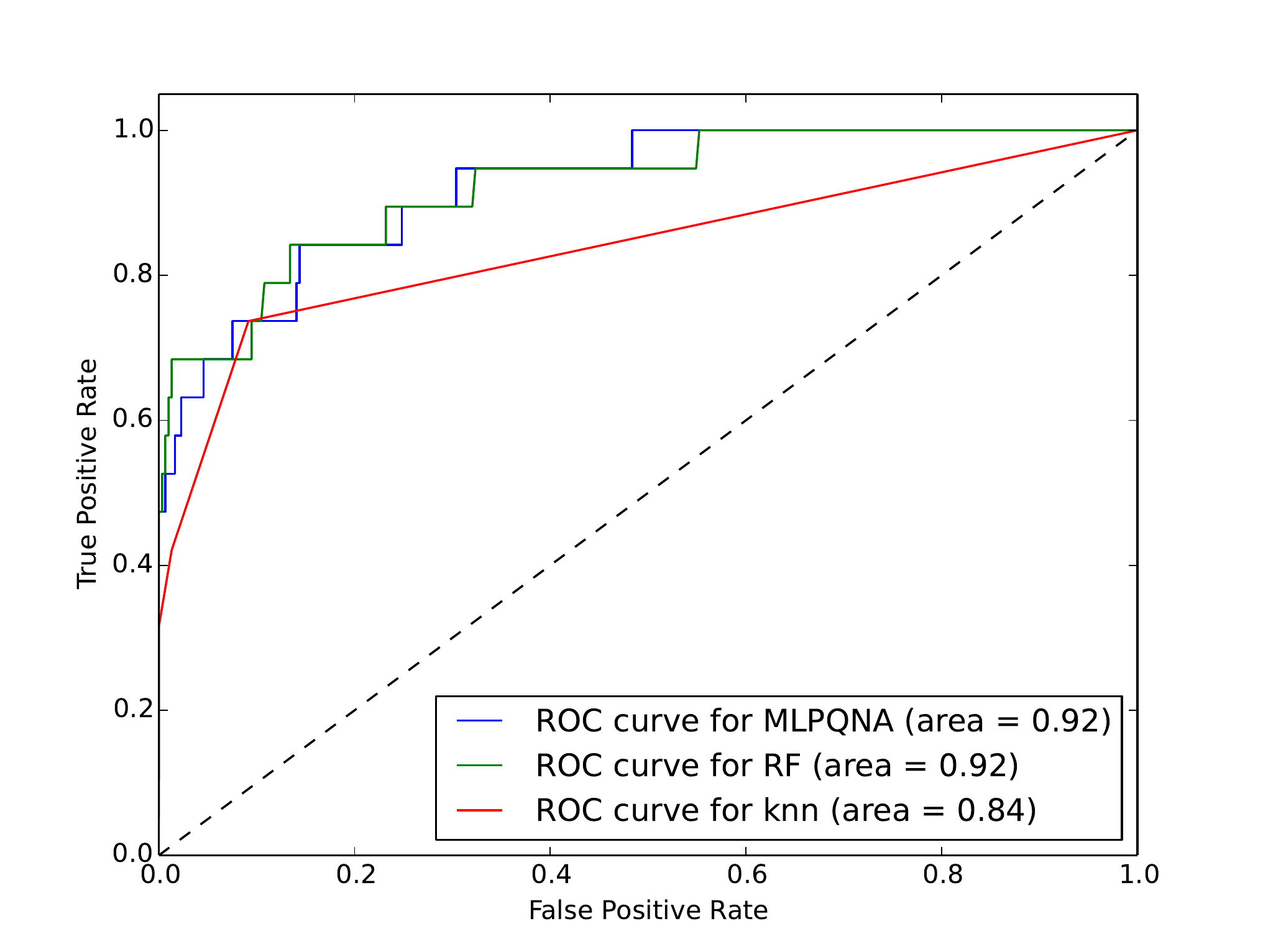}

        }
        \subfigure[\textit{RRL}]{
            \label{6class:rrl}
            \includegraphics[width=0.48\textwidth]{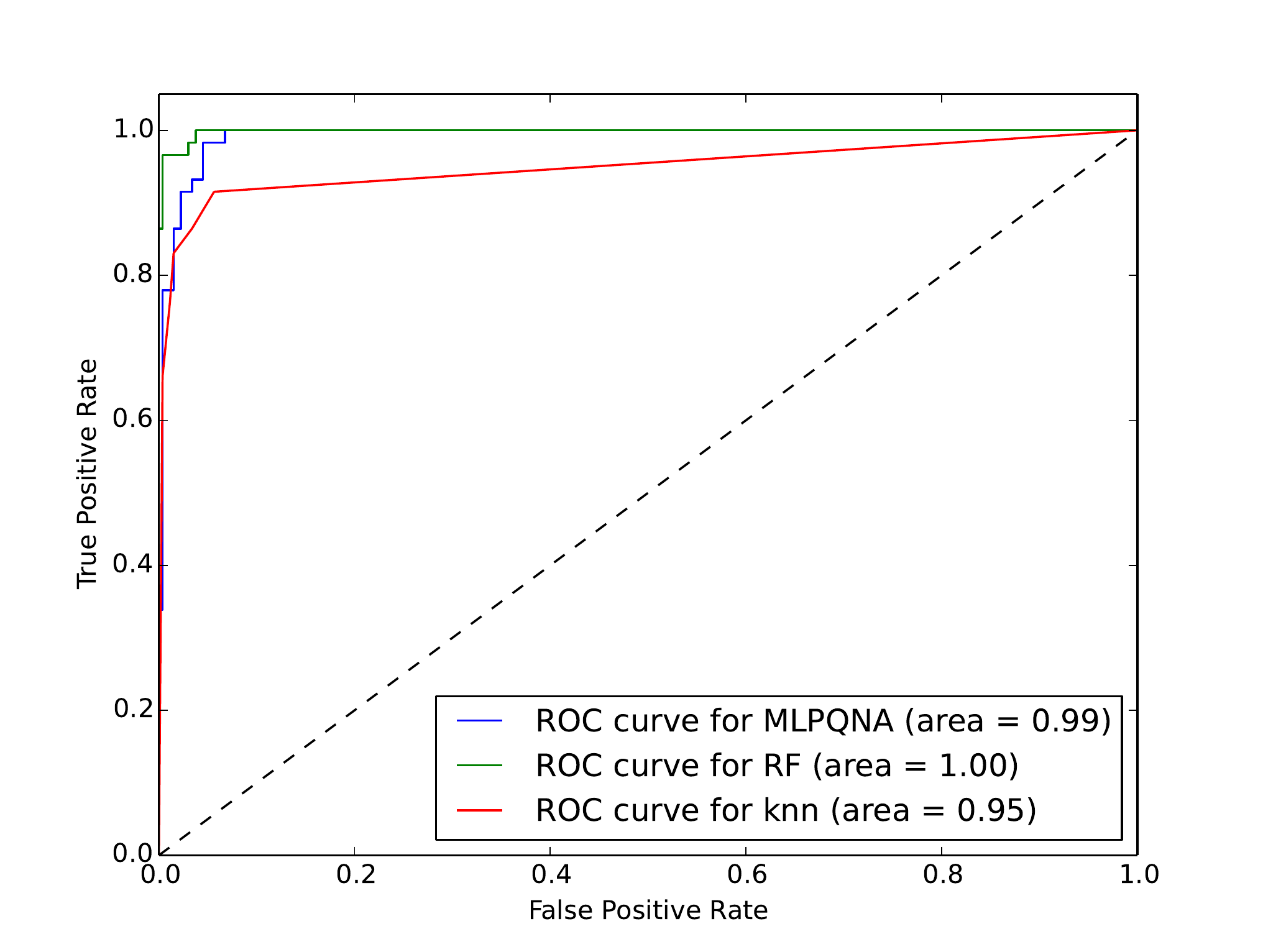}

        }

    \end{center}
    \caption{ROC curves for the six-class classification for the three models used. In the case of the KNN model the curve was obtained by taking into account the limitations imposed by the algorithm, which are determined by the choice of the number of nearest neighbors (in this case $5$ neighbors induce $20\%$ of quantization).}
   \label{roc6}
\end{figure*}

We performed the multi-class classification experiment, to understand the behavior of the classifiers in the most complex situation, i.e. considering simultaneously all the six available variable object categories.
Therefore, as explained above, we performed a preliminary experiment using the RF model with all available input features, thus obtaining the feature importance ranking for this type of classification (Fig.~\ref{list_six}). The feature ranking, in fact, is automatically provided by the RF classifier, which assigns a score to all input features, corresponding to their relevance assumed to build the decision rules of the trees during the training phase. Such information indeed is suitable to judge the weight of each individual feature in the decision process and to evaluate its eventual redundancy in terms of contribution to the learning. One useful way to exploit the feature ranking is to engage a training/test campaign, by sequentially adding features to the training parameter space (in order of their importance) and evaluating the training results, until the classification performance reaches a plateau. The final outcome of such campaign is the best compromise between the parameter space dimension and the classification performance.
After such preliminary analysis, we then submitted the dataset to the RF, MLPQNA and KNN classifiers, by using respectively all, the first $5$ and the first $3$ features of the ranking list in order of importance. A statistical evaluation of the classification results is reported in Tables \ref{cv_mlp}, \ref{cv_rf} and \ref{cv_knn}, while the ROC curves for each class are shown in Fig.~\ref{roc6}. From these results it appears evident the worst behavior of the KNN model with respect to the other classifiers. In terms of class purity, the best behavior is obtained by the RF model using all available features.

\subsection{Cataclysmic Variables vs ALL}

\begin{figure*}
\begin{center}
\includegraphics[width=15cm]{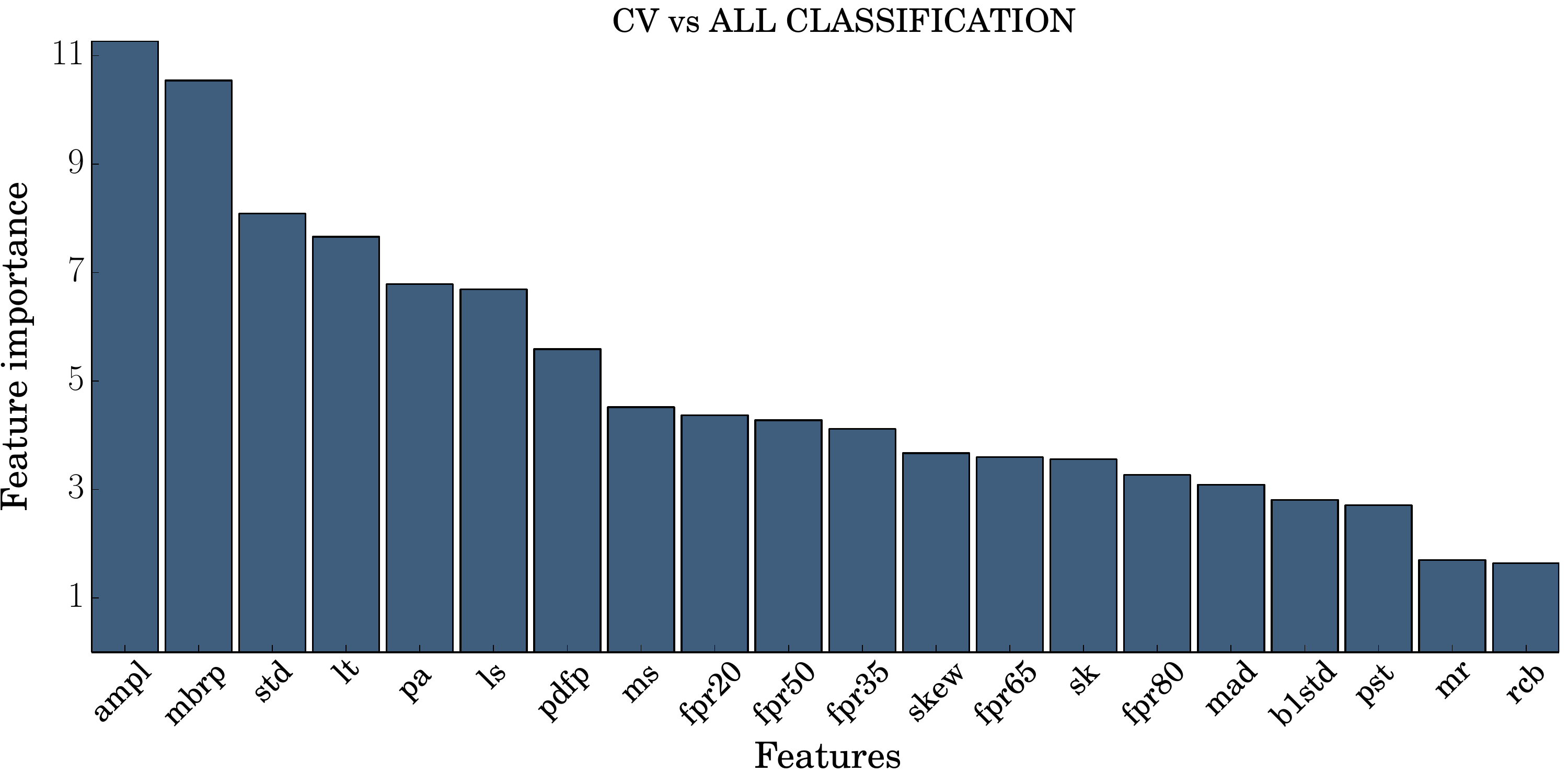}
\caption[Feature importance list obtained with RF for \textit{CV vs ALL} classification.]{Feature importance list obtained by the RF, with the importance percentage for each feature and for the \textit{CV vs ALL} classification.} \label{list_cv}
\end{center}
\end{figure*}

We started by performing an experiment using the RF model and all selected features. The data set was composed by $461$ \textit{CV} and $866$ \textit{ALL} objects.
Results are shown in Tab.~\ref{cv_rf}, while the feature ranking is given in Fig.~\ref{list_cv}.

Following the feature ranking evaluation strategy, we performed a series of experiments using the MLPQNA, RF and KNN models using different groups of features taken in order of importance: respectively, the first $3, 5, 6, 9, 10, 11$ groups and all the $20$ features listed in Fig.~\ref{list_cv}.

In most cases, groups differing by a small number of features (e.g. $5$ and $6$) led to results with similar performance and, in these cases, we retained as representative the smaller group, assuming that the most of the information is already contained into these groups.
Therefore, in the following description of experiments we explicitly report the results only for these relevant cases (see Tables \ref{cv_mlp}, \ref{cv_rf} and \ref{cv_knn}, as well as the related ROC curves in Fig.~\ref{roc}).

From this series of experiments, it appears clear that, regarding MLPQNA, the best configuration is achieved using only $5$ features after the optimization of model parameters (\textit{ampl}, \textit{mbrp}, \textit{std}, \textit{lt} and \textit{pa}), while, for the RF, the best results were obtained by retaining all $20$ features. Finally, the KNN, which is also the classifier with the worst performance, gives the best result using $6$ features only.

%%%%%%%%%%%%%%%%%%%%%
\subsection{Extra-Galactic vs Galactic}

\begin{figure*}
\begin{center}
\includegraphics[width=15cm]{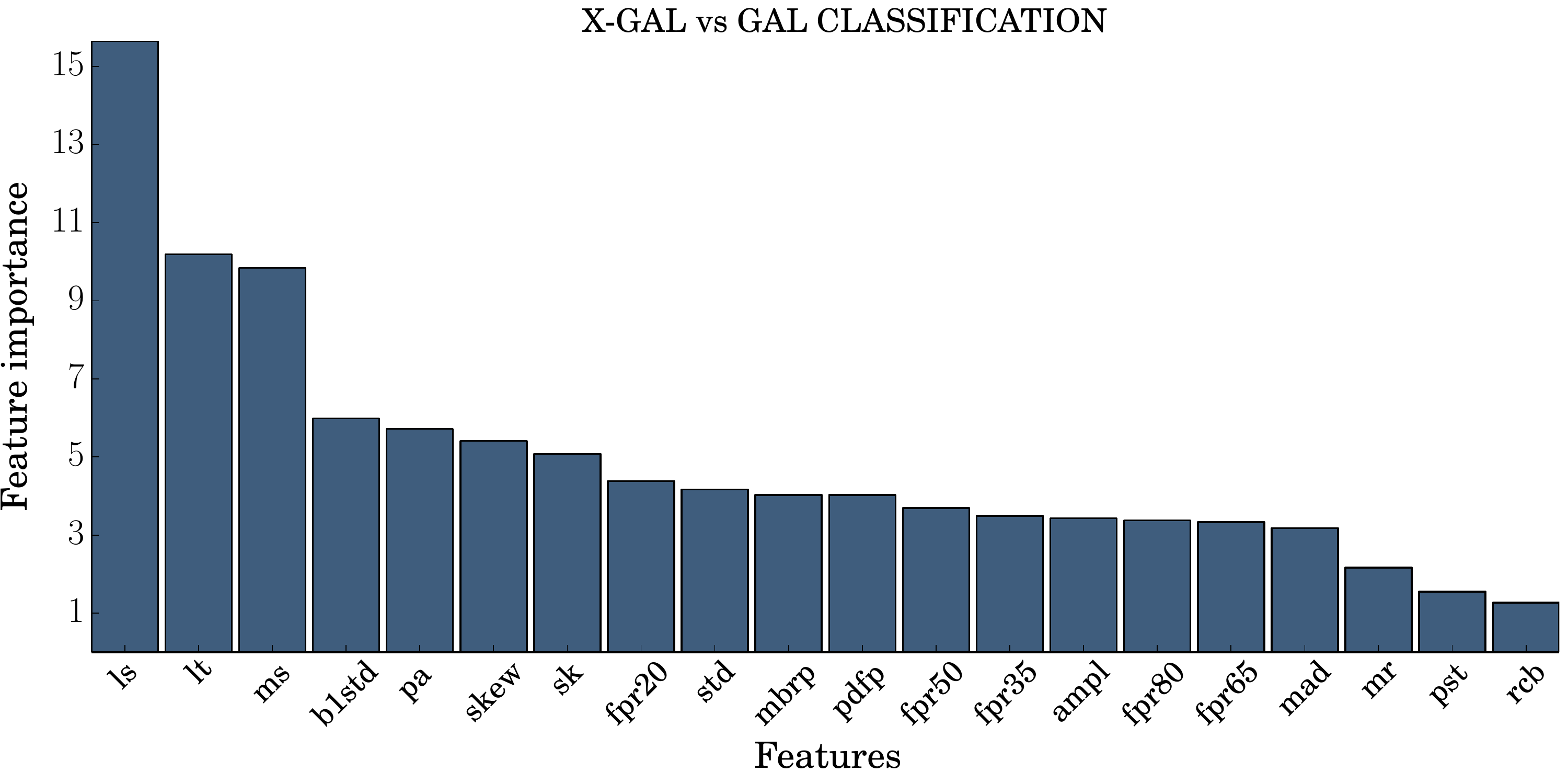}
\caption[Feature importance list obtained with RF for \textit{X-GAL vs GAL} classification.]{Feature importance list obtained by the RF, with the importance percentage for each feature and for the \textit{X-GAL vs GAL} classification.} \label{list_xgal}
\end{center}
\end{figure*}

Also in the case of the classification experiment related to $264$ EXTRA-GALACTIC, hereafter called \textit{X-GAL}, (\textit{AGN + Bl} as class $1$) patterns vs $1,063$ GALACTIC, hereafter named \textit{GAL}, (\textit{CV} + \textit{SN} + \textit{Fl} as class $2$) patterns, we first performed a feature ranking evaluation with the RF model, by using all available features (see Fig.~\ref{list_xgal}).
Again, using the ranking list and the same feature selection strategy described above, we performed a reduced number of experiments using the first $5$, $10$, and $all$ features, by applying all three ML models.

In addition, we performed one additional experiment, using the $5$ features which were selected as most relevant for the \textit{CV vs ALL} classification case. Results are presented in Tables \ref{evg_mlp}, \ref{evg_rf} and \ref{evg_knn}, while the related ROC curves are shown in Fig.~\ref{roc}. Best classification performance resulted with, respectively, $5$ features for MLPQNA (\textit{ls}, \textit{lt}, \textit{ms}, \textit{b1std} and \textit{pa}) and $10$ features for RF and KNN models (\textit{ls}, \textit{lt}, \textit{ms}, \textit{b1std}, \textit{pa}, \textit{skew}, \textit{sk}, \textit{fpr20}, \textit{std}, \textit{mbrp}).

\begin{figure*}
\begin{center}
\includegraphics[width=15cm]{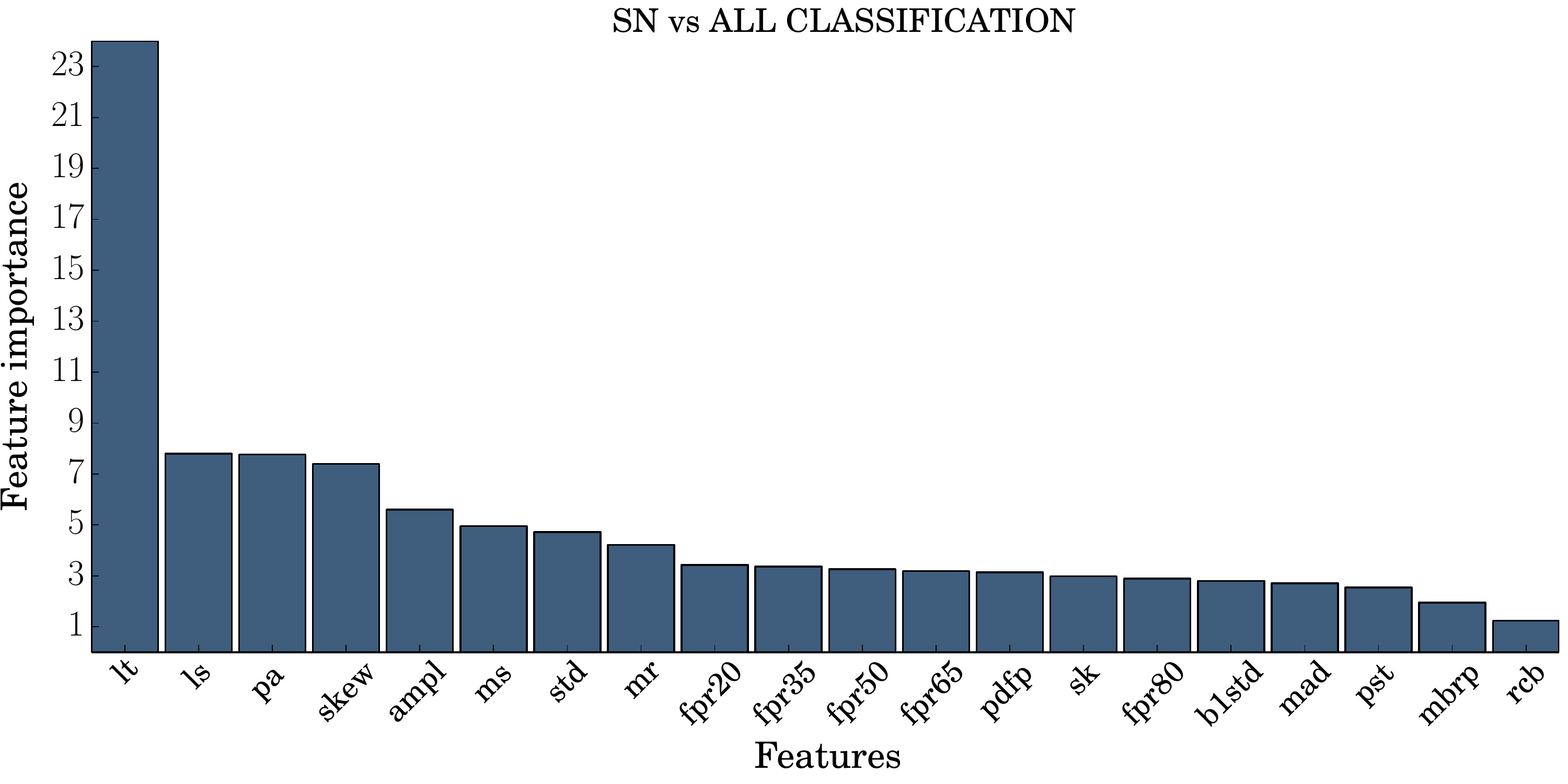}
\caption[Feature importance list obtained with RF for \textit{SN vs ALL} classification.]{Feature importance list obtained by the RF, with the importance percentage for each feature and for the \textit{SN vs ALL} classification.} \label{list_sn}
\end{center}
\end{figure*}

\subsection{Supernovae vs ALL}
Finally, we performed experiments for \textit{Supernovae} (\mbox{class $1$}), versus \textit{ALL} (all other classes, labeled as class $2$), but in this case we added to the second group also the sixth class containing \textit{RR Lyrae}, thus obtaining a sample of $536$ \textit{SN} and $1,083$ \textit{ALL} class objects. Again, we started from the feature importance evaluation shown in Fig.~\ref{list_sn}.

\begin{figure*}
     \begin{center}

        \subfigure[\textit{CV vs ALL}]{
            \label{venn:cv}
            \includegraphics[width=0.50\textwidth]{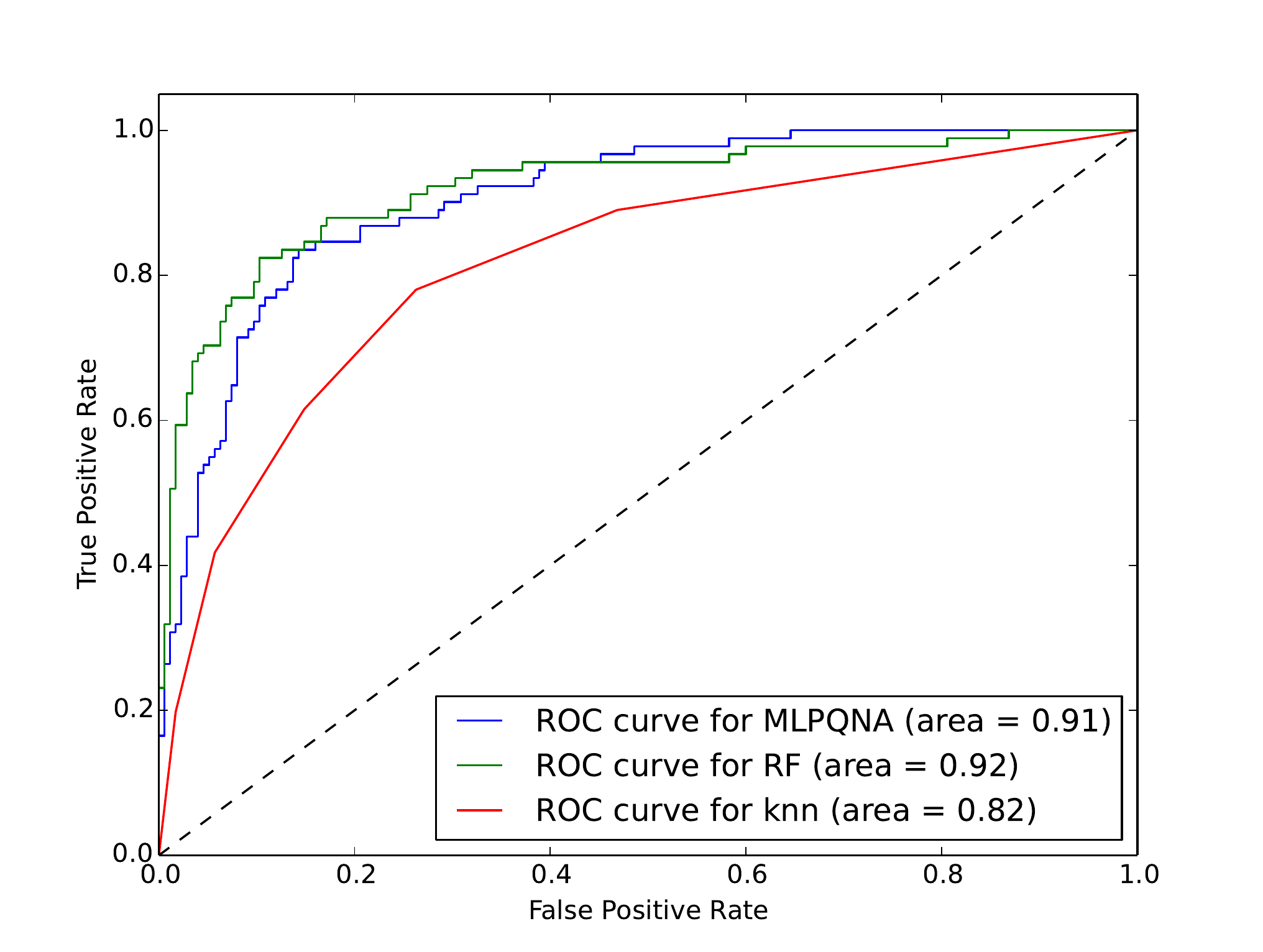}

        }\\
        \subfigure[\textit{X-GAL vs GAL}]{
           \label{venn:evg}
           \includegraphics[width=0.50\textwidth]{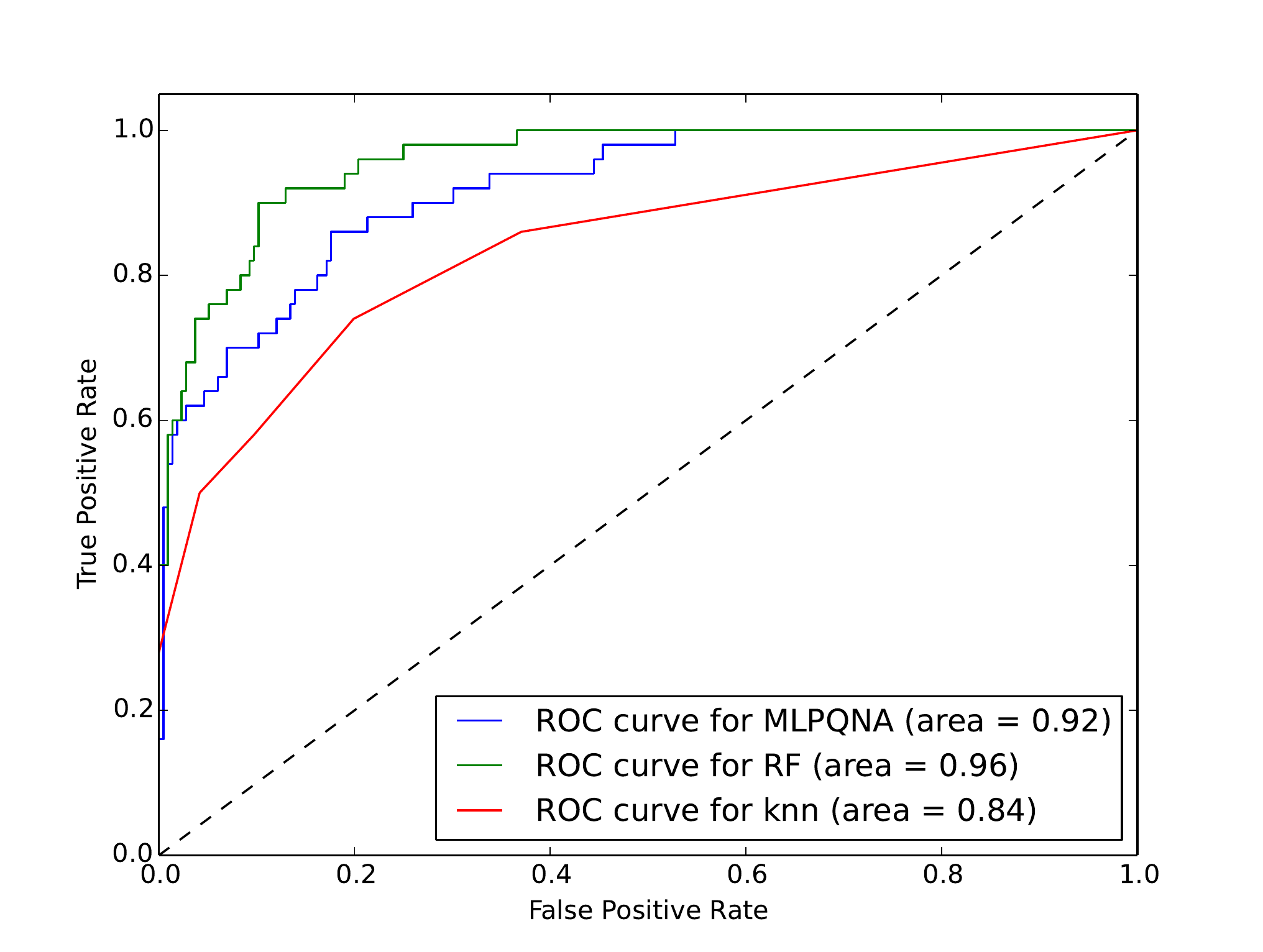}

        }\\
        \subfigure[\textit{SN vs ALL}]{
            \label{venn:sn}
            \includegraphics[width=0.50\textwidth]{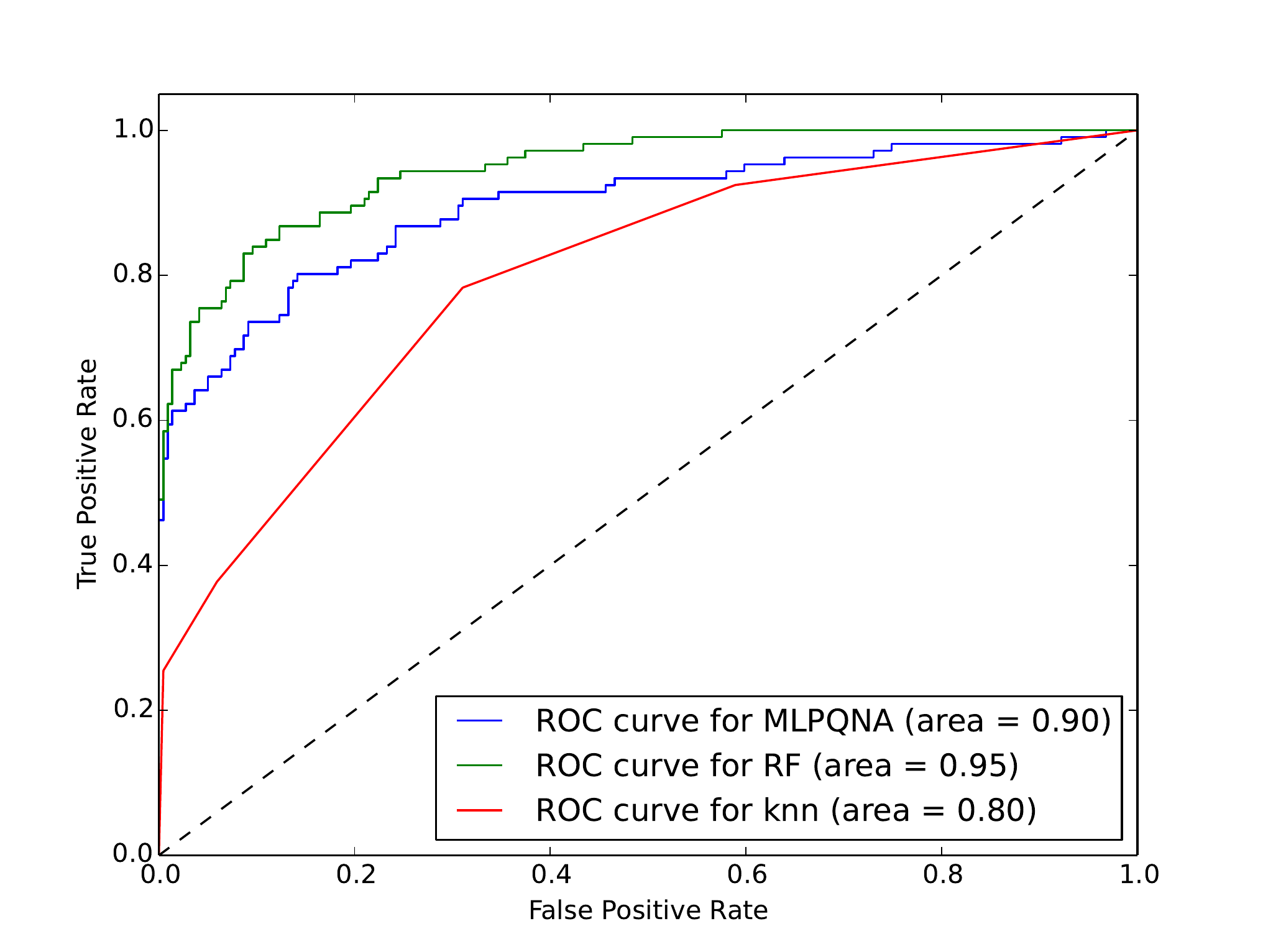}

        }

    \end{center}
    \caption{ROC curves for the three different types of classification in the three experiment types. n the case of the KNN model the curve was obtained by taking into account the limitations imposed by the algorithm, which are determined by the choice of the number of nearest neighbors (in this case $5$ neighbors induce $20\%$ of quantization).}
   \label{roc}
\end{figure*}

As it was already done in the previous cases, we performed the classification experiments with the RF, MLPQNA and KNN models. We report here the results obtained in the cases of, respectively, the first $3$, $5$ and $10$ features in the ranking list. Moreover, we performed additional experiments using the best group of $5$ features obtained from the \textit{CV vs ALL} experiment (see Fig.~\ref{list_cv}).
Results for the three experiments are reported in Tables \ref{sn_mlp}, \ref{sn_rf}, \ref{sn_knn} and ROC curves in Fig.~\ref{roc}. The best classification performance have been obtained with, respectively, $10$ features for RF model (\textit{lt}, \textit{ls}, \textit{pa}, \textit{skew}, \textit{ampl}, \textit{ms}, \textit{std}, \textit{mr}, \textit{fpr20}, \textit{fpr35}) and only $3$ features for MLPQNA and KNN classifiers (\textit{lt}, \textit{ls}, \textit{pa}).

\section{Discussion}\label{SEC:discussion}

From the experiments previously described, we can notice that, in this context (as imposed by the structure of the parameter space and the size of the data), the Random Forest performs on average slightly better than MLPQNA and objectively better than KNN.

The results presented in the previous paragraph show that at least in presence of such a limited training set the \textit{six-class} experiment is outperformed by the binary classification experiments. The performance achieved by the RF and MLPQNA models for the classes which are more relevant for our work, for instance SNs and CVs categories, led us to investigate two cases of binary classification, respectively, \textit{SN vs ALL} and \textit{CV vs ALL}. Furthermore, we approached also the possibility to enclose Blazars and AGN in a single class compared with other categories, thus obtaining a third binary classification experiment, named \textit{X-GAL vs GAL}. We removed the RR Lyrae category from the binary classification experiments, due to their periodic behavior, which introduces a very well defined signature in the data. This has been also derived from the multi-class experiment results, showing how the \textit{RR Lyrae} objects are easy to classify, thus being not required their inclusion. Only in the case of the \textit{SN vs ALL} experiment, in order to be as general as possible, we re-introduced the RR Lyrae category.

A first interesting result is that, in spite of the ranking orders obtained for the different experiments and of the results assigned as best, in all cases an accuracy above $80\%$ of efficiency is obtained using the same $5$ most relevant features of the experiment \textit{CV vs ALL} (\textit{ampl}, \textit{mbrp}, \textit{std}, \textit{lt} and \textit{pa}).
This can be understood by comparing the first five positions of the ranking list obtained from the RF for all classification cases, as reported in Figures \ref{list_cv}, \ref{list_xgal} and \ref{list_sn}.
In fact, we can notice that among the first five features of Fig.~\ref{list_cv}, there are two (\textit{lt} and \textit{pa}) in common with other cases,
while the two features \textit{ampl} and \textit{ls} are in common between two groups of features (Figures \ref{list_cv} and \ref{list_sn}). Moreover, the feature \textit{std} is often present within the best groups among different experiments.

Concerning the MCC, this value is almost always above $0.50$ for the MLPQNA and RF. In fact, just one experiment shows an MCC below this value, while the best one is $0.74$. Therefore, we can conclude that the observed classification with these three classifiers, is close to the expected one, and that the model shows a proper behavior.

The three classifiers perform differently on different types of objects and, as usual in classification experiments, this implies that the overall performance can be increased by combining the output of the three models. To verify this hypothesis we analyzed the overall efficiency variation by taking into account the objects classified by single models and those equally classified by the combination of MLPQNA and RF, MLPQNA and KNN, RF and KNN, and by all three classifiers together.

\begin{figure*}
     \begin{center}

        \subfigure[\textit{CV vs ALL}]{
            \label{venn:cv}
            \includegraphics[width=0.50\textwidth]{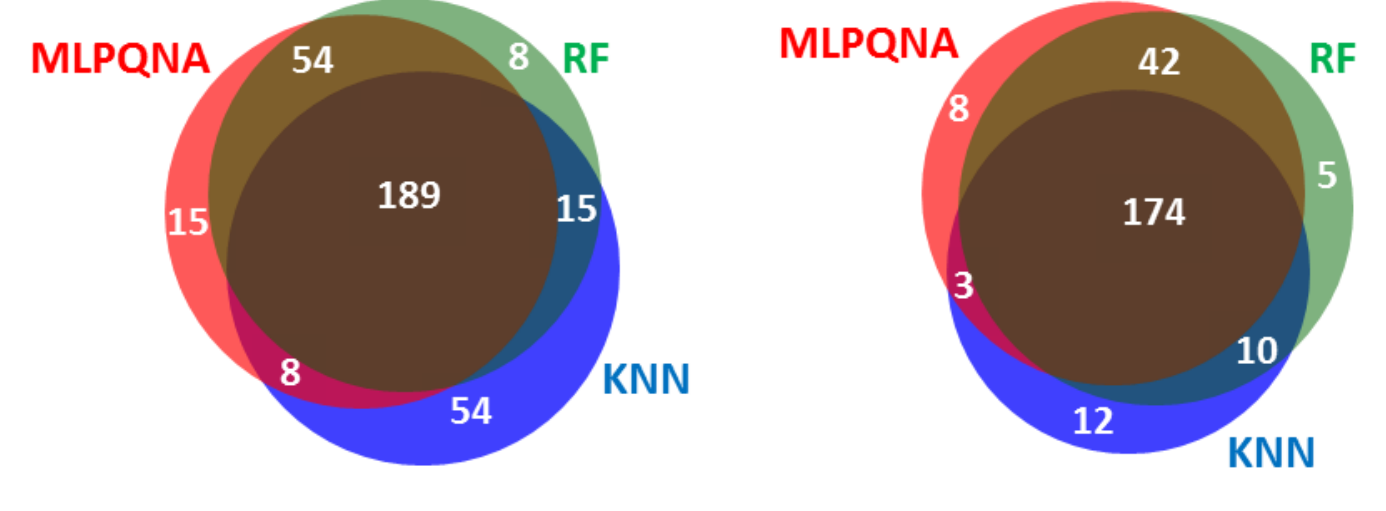}

        }\\
        \subfigure[\textit{X-GAL vs GAL}]{
           \label{venn:evg}
           \includegraphics[width=0.50\textwidth]{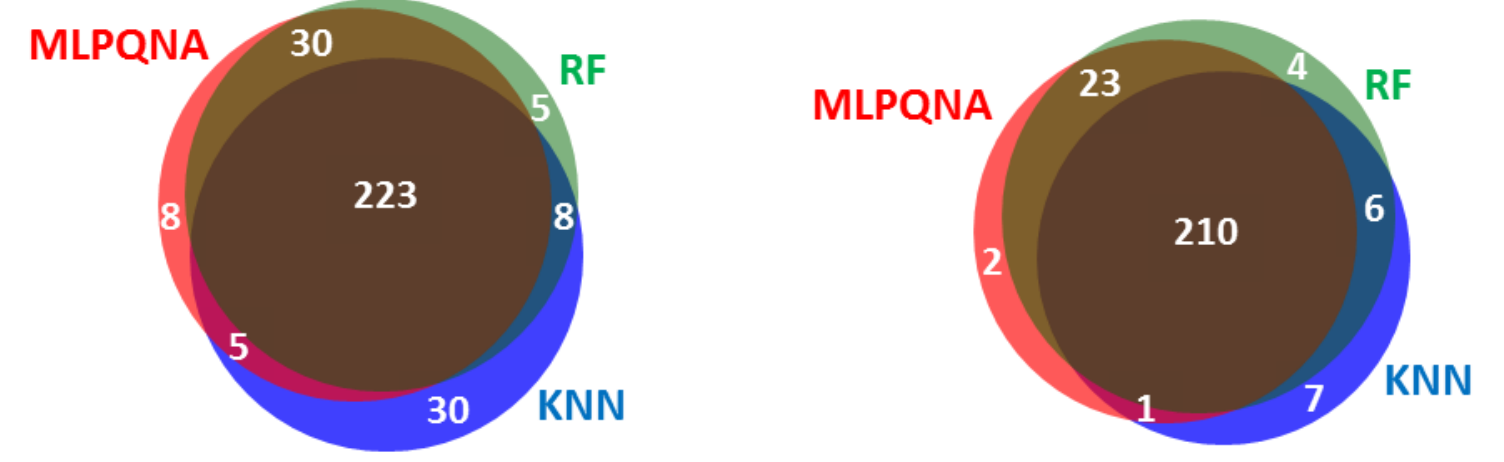}

        }\\
        \subfigure[\textit{SN vs ALL}]{%
            \label{venn:sn}
            \includegraphics[width=0.50\textwidth]{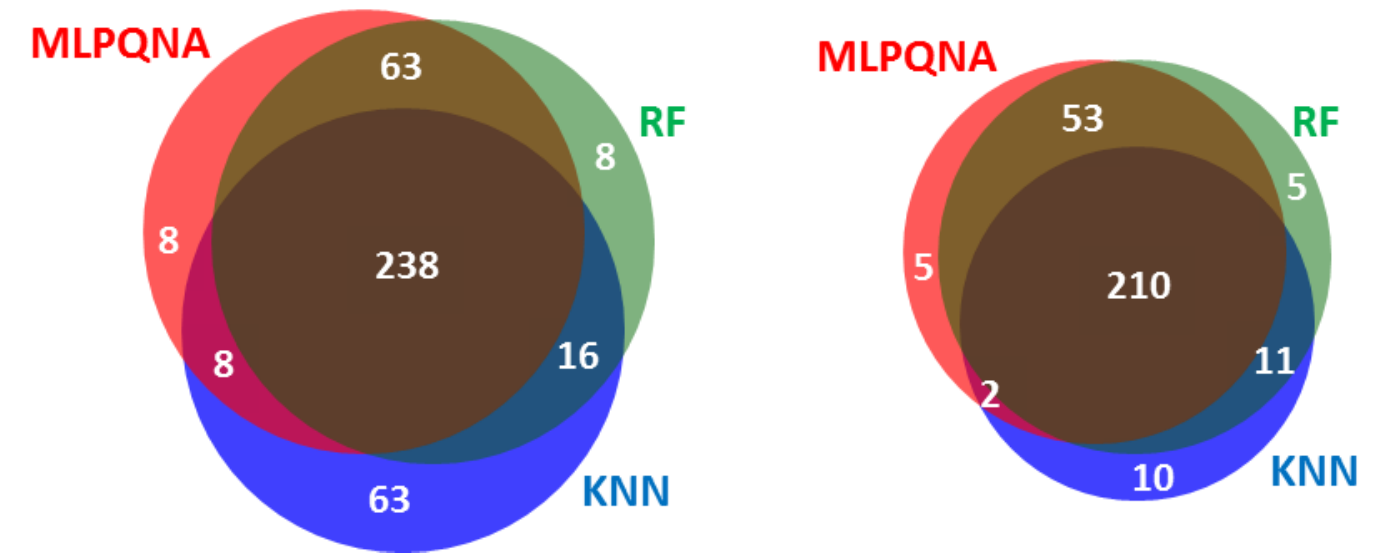}

        }%

    \end{center}
    \caption{Venn diagrams showing all the objects (left column) and the correctly classified objects (right column), based on efficiency, for the three different types of classification in the three experiment types. The intersection areas then show the objects that are classified in the same way by different methods. Values are taken from Tables \ref{cv_stat}, \ref{evg_stat} and \ref{sn_stat} respectively.}
   \label{venn}
\end{figure*}

For this analysis, shown in Tables \ref{cv_stat}, \ref{evg_stat} and \ref{sn_stat}, we performed experiments by randomly splitting the catalogue into a training and a blind test set, containing respectively the $80\%$ and the $20\%$ of the data. The increase in performance is quite evident. These results are also visualized as Venn-diagrams in Fig.~\ref{venn}.

\begin{table}\scriptsize
 \centering
 \begin{tabular}{| c | c | c |}
 \hline
 \textbf{\textit{CV vs ALL}}&\textbf{Size}&\textbf{Fraction}\\
 \hline
 \hline
 \textit{Total test objects}&266&-\\
 \hline
 \textit{MLPQNA Eff}&224&84\%\\
 \hline
 \textit{RF Eff}&231&87\%\\
 \hline
 \textit{KNN Eff}&199&75\%\\
 \hline
 \textit{(MLPQNA \& RF \& KNN) equally classified}&189&71\%\\
 \hline
 \textit{(MLPQNA \& RF) Eff}&216&89\%\\
 \hline
 \textit{(MLPQNA \& KNN) Eff}&177&90\%\\
 \hline
 \textit{(RF \& KNN) Eff}&184&90\%\\
 \hline
 \textit{(MLPQNA \& RF \& KNN) Eff}&174&92\%\\
 \hline
 \end{tabular}
 \caption{Statistical analysis on the test output for the best experiments of \textit{CV vs ALL} classification for the three models ($5*$ in Tab.~\ref{cv_mlp} for the MLPQNA, $20$ in Tab.~\ref{cv_rf} for the RF, and $6$ in Tab.~\ref{cv_knn} for the KNN). The first row reports the total amount of test objects. Second, third and fourth rows indicate the overall efficiency obtained by the three models. While the fifth row reports the number of objects equally classified by the three models (i.e. only the objects for which the three models provide the same classification). Finally the last four rows report the overall efficiencies referred only to the equally classified objects.}
 \label{cv_stat}
 \end{table}

 \begin{table}\scriptsize
 \centering
 \begin{tabular}{| c | c | c |}
 \hline
 \textbf{X-GAL vs GAL}&\textbf{Size}&\textbf{Fraction}\\
 \hline
 \hline
 \textit{Total test objects}&266&-\\
  \hline
 \textit{MLPQNA Eff}&236&89\%\\
 \hline
 \textit{RF Eff}&243&91\%\\
 \hline
 \textit{KNN Eff}&224&84\%\\
  \hline
 \textit{(MLPQNA \& RF \& KNN) equally classified}&223&84\%\\
 \hline
 \textit{(MLPQNA \& RF) Eff}&233&92\%\\
 \hline
 \textit{(MLPQNA \& KNN) Eff}&211&92\%\\
 \hline
 \textit{(RF \& KNN) Eff}&216&93\%\\
 \hline
 \textit{(MLPQNA \& RF \& KNN) Eff}&210&94\%\\
 \hline
 \end{tabular}
 \caption{Statistical analysis on the test output for the best experiments of \textit{X-GAL vs GAL} classification for the three models ($5*$ in Tab.~\ref{evg_mlp} for the MLPQNA, $10$ in Tab.~\ref{evg_rf} for the RF, and $10$ in Tab.~\ref{evg_knn} for the KNN). The first row reports the total amount of test objects. Second, third and fourth rows indicate the overall efficiency obtained by the three models. While the fifth row reports the number of objects equally classified by the three models (i.e. only the objects for which the three models provide the same classification). Finally the last four rows report the overall efficiencies referred only to the equally classified objects.}
 \label{evg_stat}
 \end{table}

 \begin{table}\scriptsize
 \centering
 \begin{tabular}{| c | c | c |}
 \hline
 \textbf{\textit{SN vs ALL}}&\textbf{Size}&\textbf{Fraction}\\
 \hline
 \hline
 \textit{Total test objects}&325&-\\
 \hline
 \textit{MLPQNA Eff}&278&85\%\\
 \hline
 \textit{RF Eff}&288&89\%\\
 \hline
 \textit{KNN Eff}&241&74\%\\
 \hline
 \textit{(MLPQNA \& RF \& KNN) equally classified}&238&73\%\\
 \hline
 \textit{(MLPQNA \& RF) Eff}&271&90\%\\
 \hline
 \textit{(MLPQNA \& KNN) Eff}&220&89\%\\
 \hline
 \textit{(RF \& KNN) Eff}&229&90\%\\
 \hline
 \textit{(MLPQNA \& RF \& KNN) Eff}&218&91\%\\
 \hline
 \end{tabular}
 \caption{Statistical analysis on the test output for the best experiments of \textit{SN vs ALL} classification for the three models ($3*$ in Tab.~\ref{sn_mlp} for the MLPQNA, $10$ in Tab.~\ref{sn_rf} for the RF, and $3$ in Tab.~\ref{sn_knn} for the KNN). The first row reports the total amount of test objects. Second, third and fourth rows indicate the overall efficiency obtained by the three models. While the fifth row reports the number of objects equally classified by the three models (i.e. only the objects for which the three models provide the same classification). Finally the last four rows report the overall efficiencies referred only to the equally classified objects.}
 \label{sn_stat}
 \end{table}

\begin{figure*}
\begin{center}

\subfigure[]{
            \includegraphics[width=0.48\textwidth]{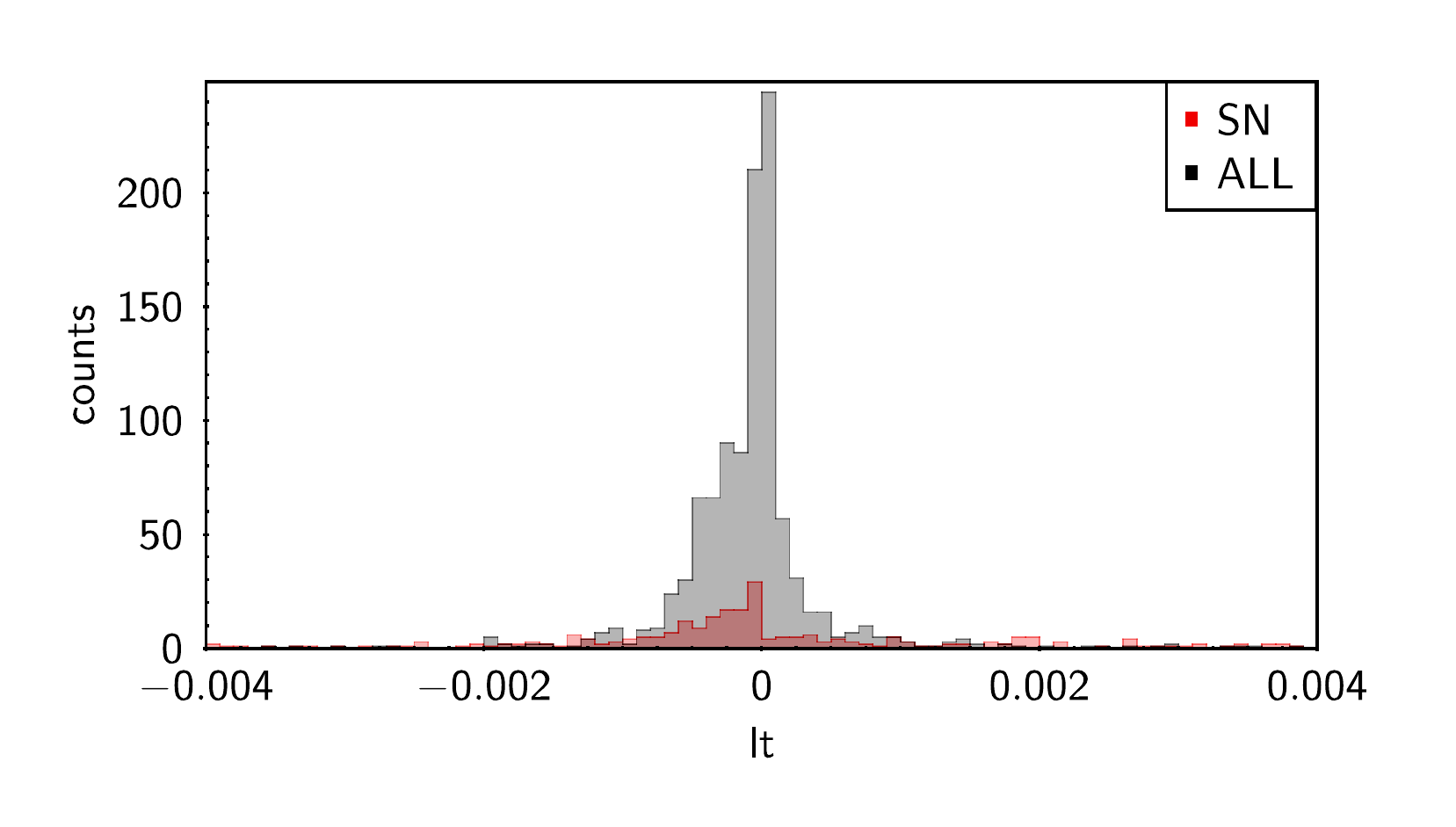}
        }
\subfigure[]{
            \includegraphics[width=0.48\textwidth]{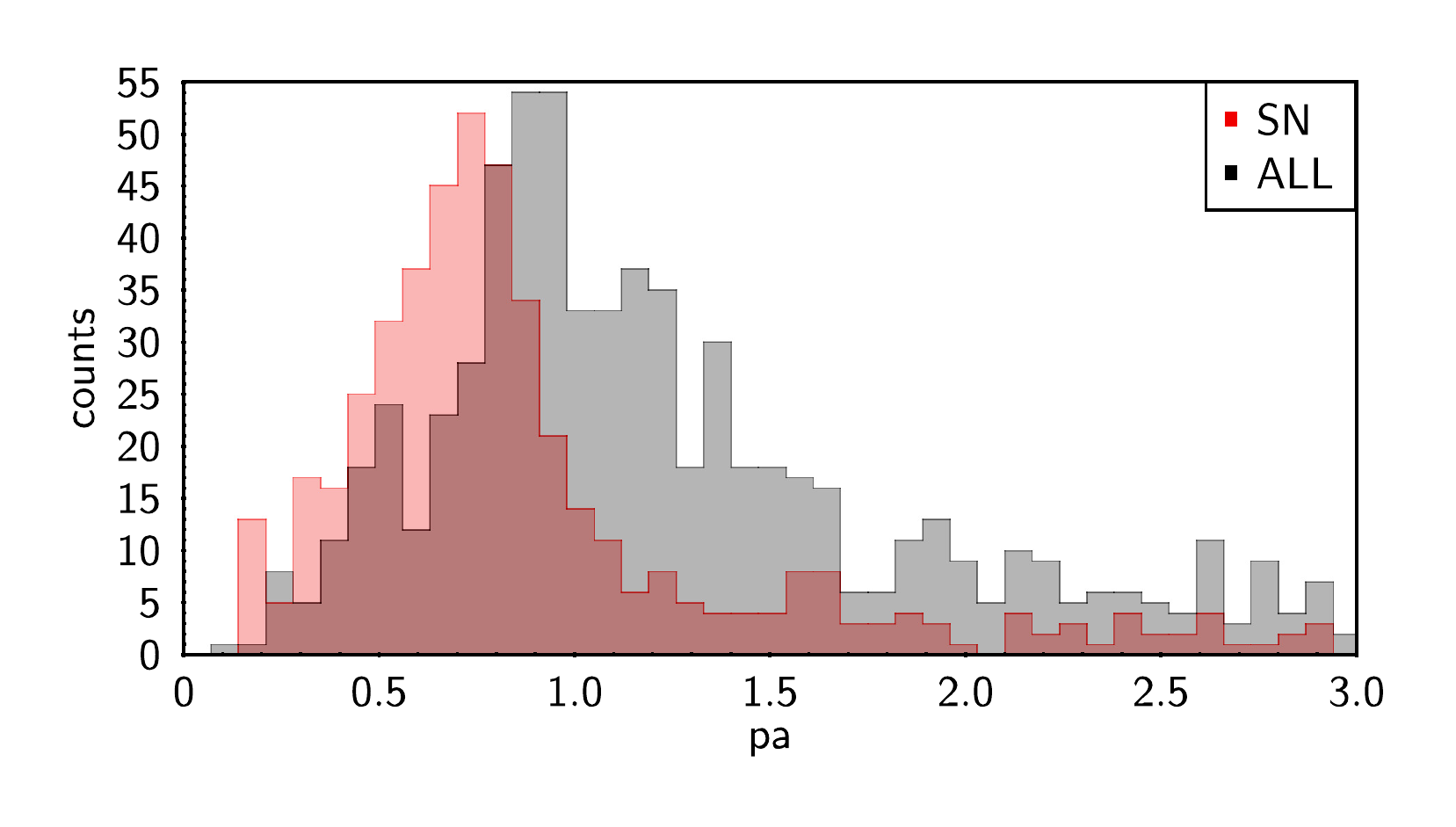}
        }\\
\subfigure[]{
            \includegraphics[width=0.48\textwidth]{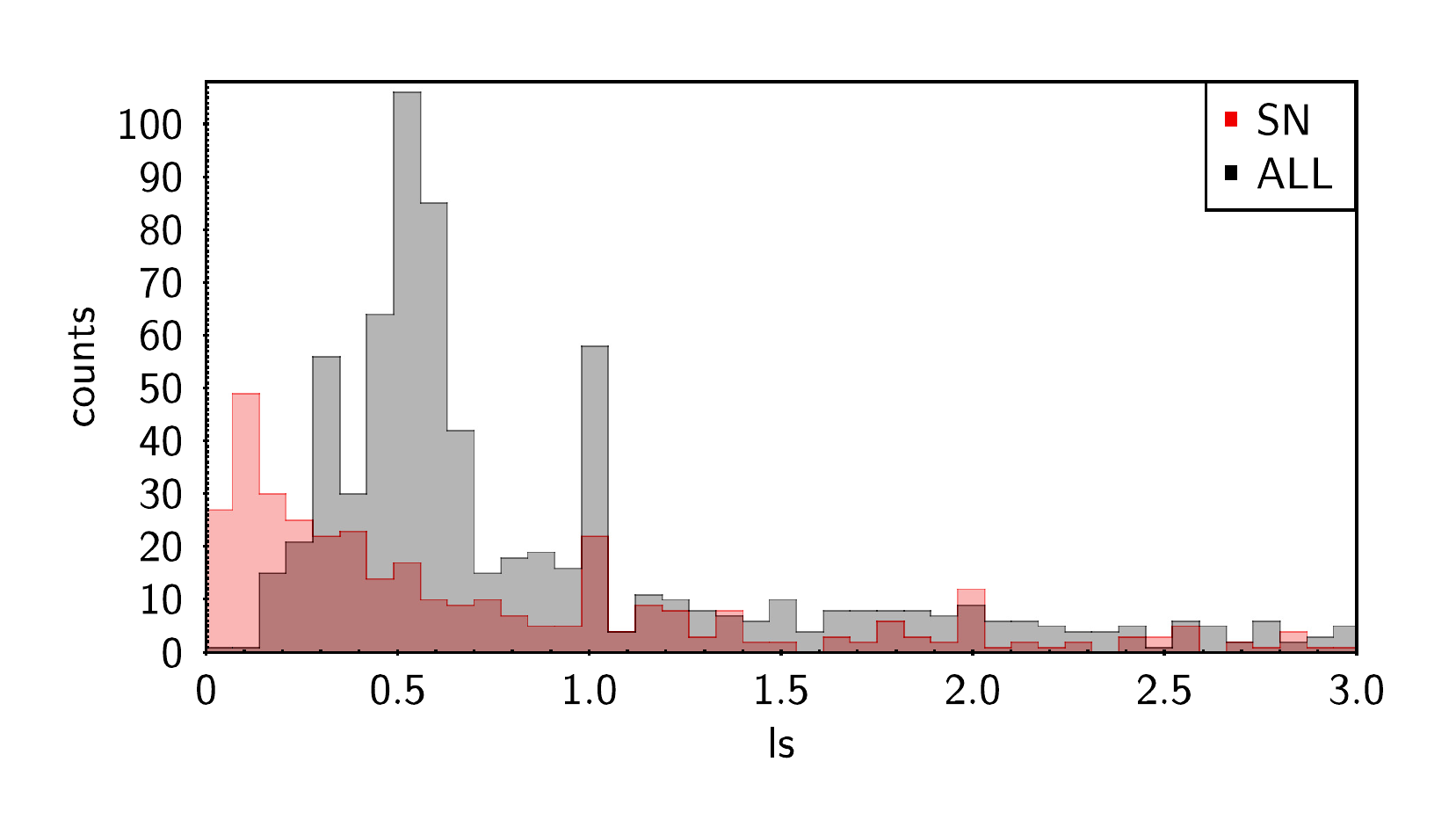}
        }
\subfigure[]{
            \includegraphics[width=0.48\textwidth]{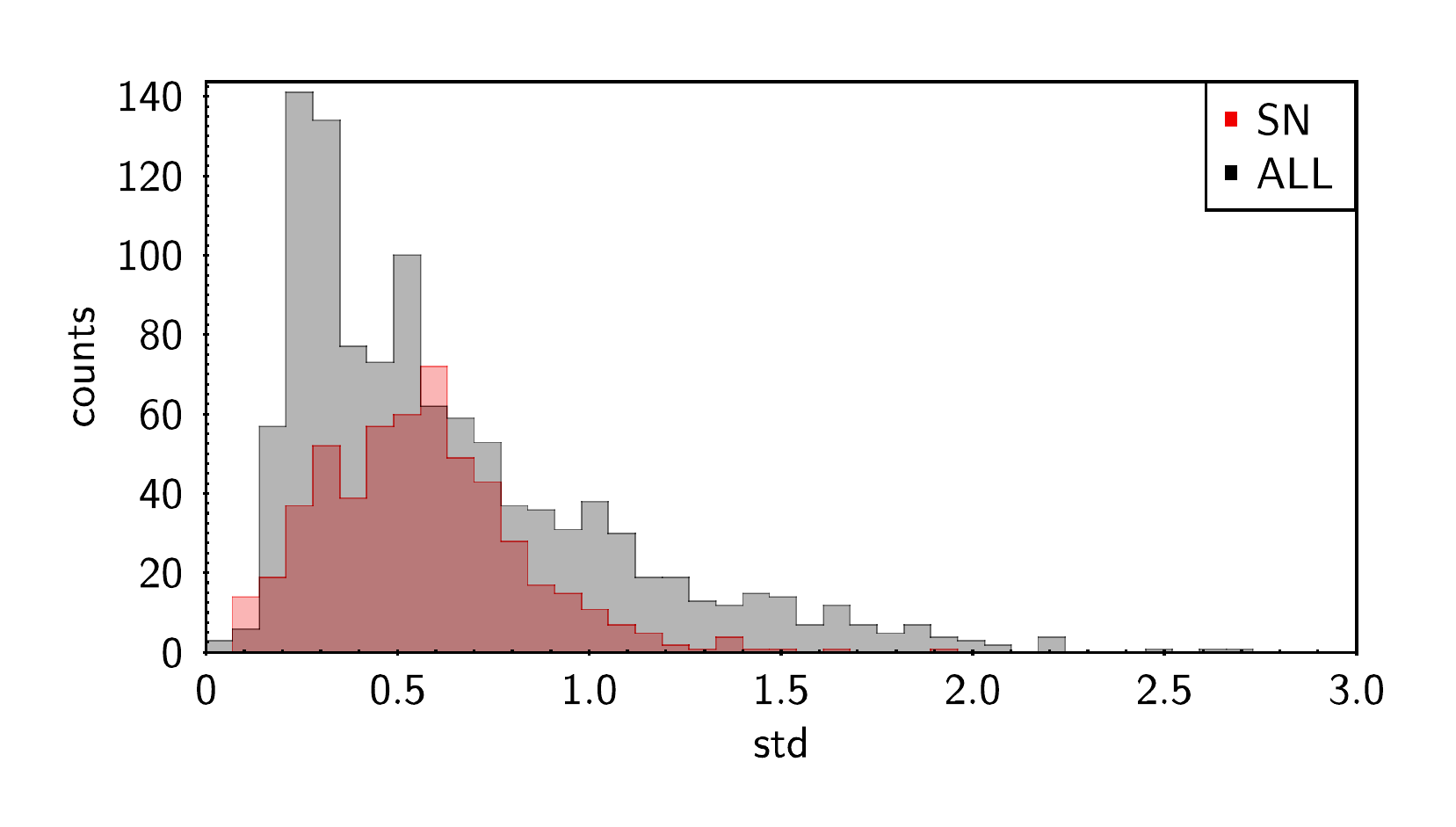}
        }\\
\subfigure[]{
            \includegraphics[width=0.48\textwidth]{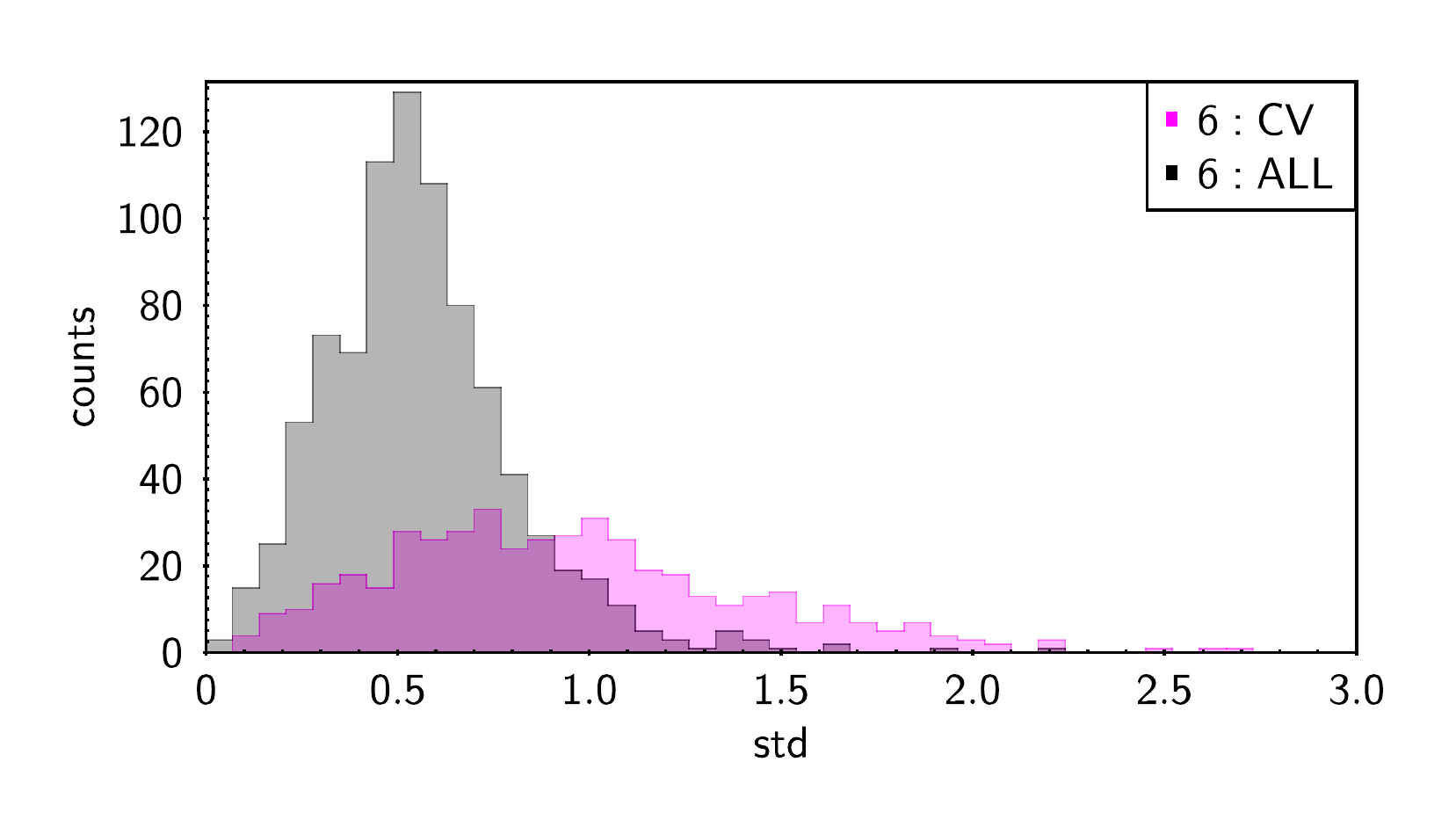}
        }
\subfigure[]{
            \includegraphics[width=0.48\textwidth]{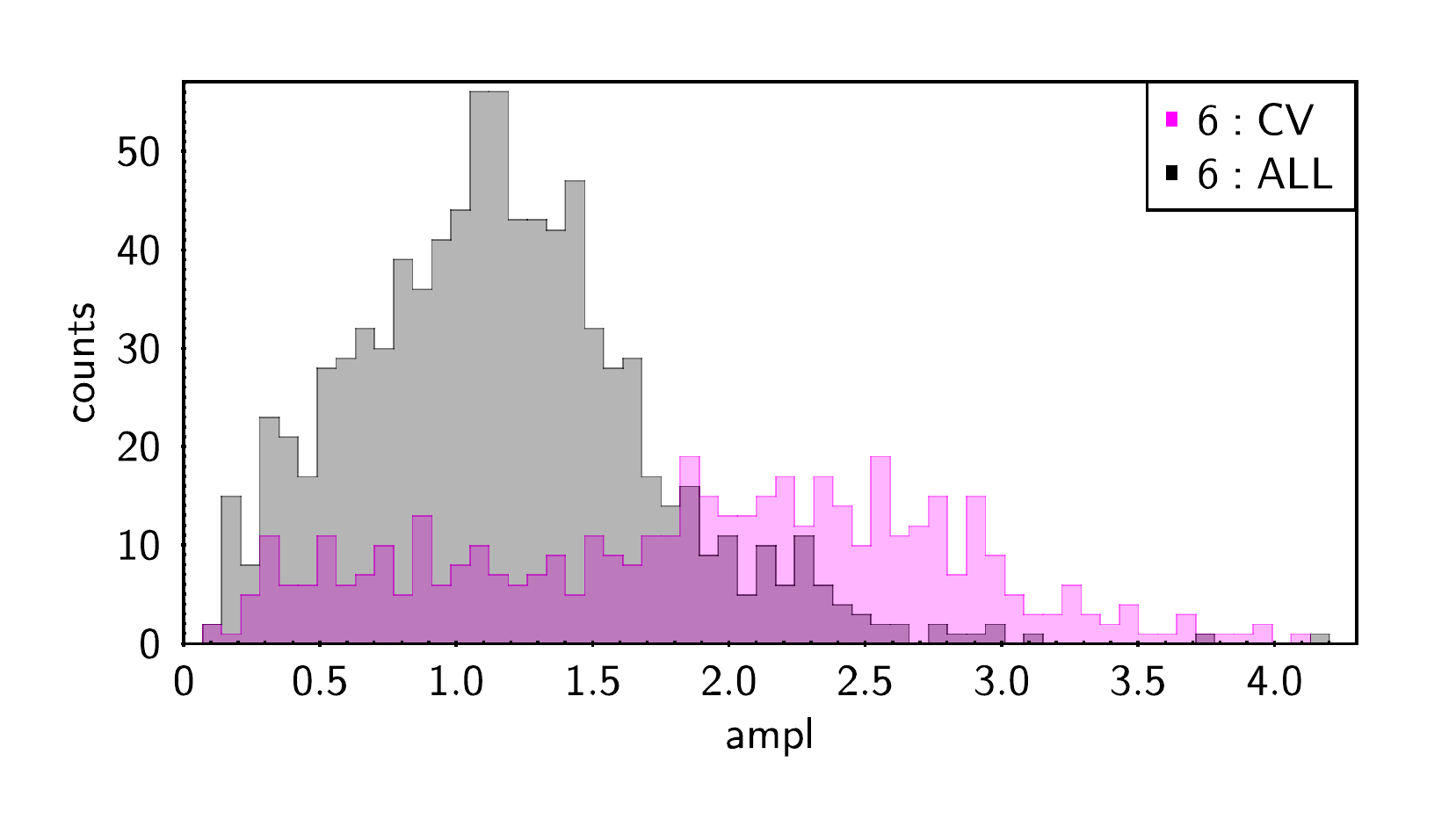}
        }\\
\end{center}
\caption{Distribution of the \textit{lt} (panel a), \textit{pa} (panel b), \textit{ls} (panel c) and \textit{std} (panel d) in the case \textit{SN vs ALL} experiment. The diagram shows a zoomed portion of the distribution to better visualize the region of interest. Red color is related to \textit{SN} objects, dark gray color to \textit{ALL} class objects, while dark brown shows the overlay area of the histogram. Panels (e) and (f): distribution of the, respectively, \textit{std} and \textit{ampl} features in the case \textit{CV vs ALL} experiment. Purple color is related to \textit{CV} objects, dark gray represent the \textit{ALL} class objects, while in dark purple is shown the overlay area of the histogram.}
\label{cv_ampl}
\end{figure*}

The relevance of the various features in the experiments can be better investigated by looking at their distributions. For the sake of clarity in Fig.~\ref{cv_ampl} we show a few relevant examples.
In panels a, b and c we show the distribution of the features \textit{lt}, \textit{pa} and \textit{ls} for the \textit{SN vs ALL}
experiment while in panel d and e, we show instead the distribution the parameter \textit{std} in the \textit{SN versus ALL} and in the
\textit{CV vs ALL} experiments. Finally, in panel f, we show the distribution of the \textit{ampl} feature in the \textit{CV vs ALL} experiment.
In all cases, what appears evident is that individual features fail to separate unequivocally the classes, thus confirming that their combination is needed to achieve a proper classification. Nevertheless the different roles played by the \textit{std} (panels d and e) in the experiments \textit{SN vs ALL} and \textit{CV vs ALL} (cf. figures \ref{list_sn} and \ref{list_cv}, respectively) is confirmed by the histograms.

Given the peculiar shape of the \textit{SN} light curves, it is not a surprise that in the experiment \textit{SN vs ALL},
the \textit{lt} has a relevance of $24\%$ followed in third position by \textit{pa} with a relevance of $7.7\%$.
The fact that in this experiment the Lomb-Scargle index (\textit{ls}) is ranked second, might seem strange
since it is used as an indication of periodic behavior. The histogram in panel c shows, however, that this
is due to the fact that on average objects in the \textit{SN} class (being non periodic) have a \textit{ls} much
smaller than the \textit{ALL} class.

%More difficult is to understand why the parameter lt (linear trend) which carries the $24\%$ of the information
%in the \textit{SN vs ALL} experiment has a distribution which is quite similar in both classes.

In the specific context of the CRTS, a completeness of $\sim96\%$ and a purity of $84\%$ in the \textit{SN vs ALL} classification experiment imply that the sample of candidate \textit{SNs} produced with our method, would correctly identify $\sim2520$ out of the $2631$ confirmed \textit{SNs} and would produce a sample of $\sim420$ possibly spurious objects. These results, however cannot be easily extrapolated to other surveys, since the performance of the method depends drastically on the parameter space covered by the training sample, which as it has been discussed before, is strictly depending on the specific survey.

\begin{figure*}
\begin{center}
\subfigure[]{
            \includegraphics[width=0.80\textwidth]{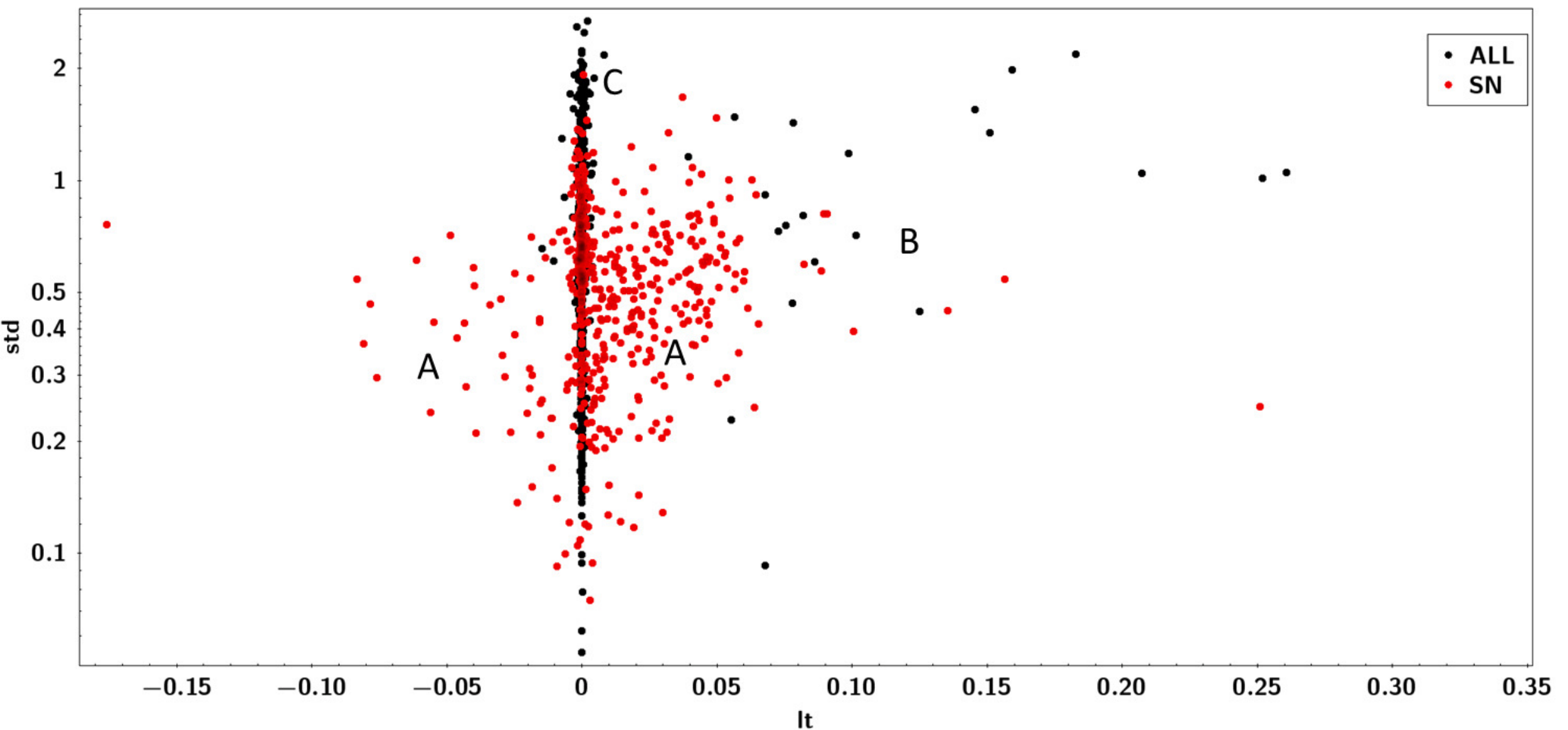}
        }\\
\subfigure[]{
            \includegraphics[width=0.80\textwidth]{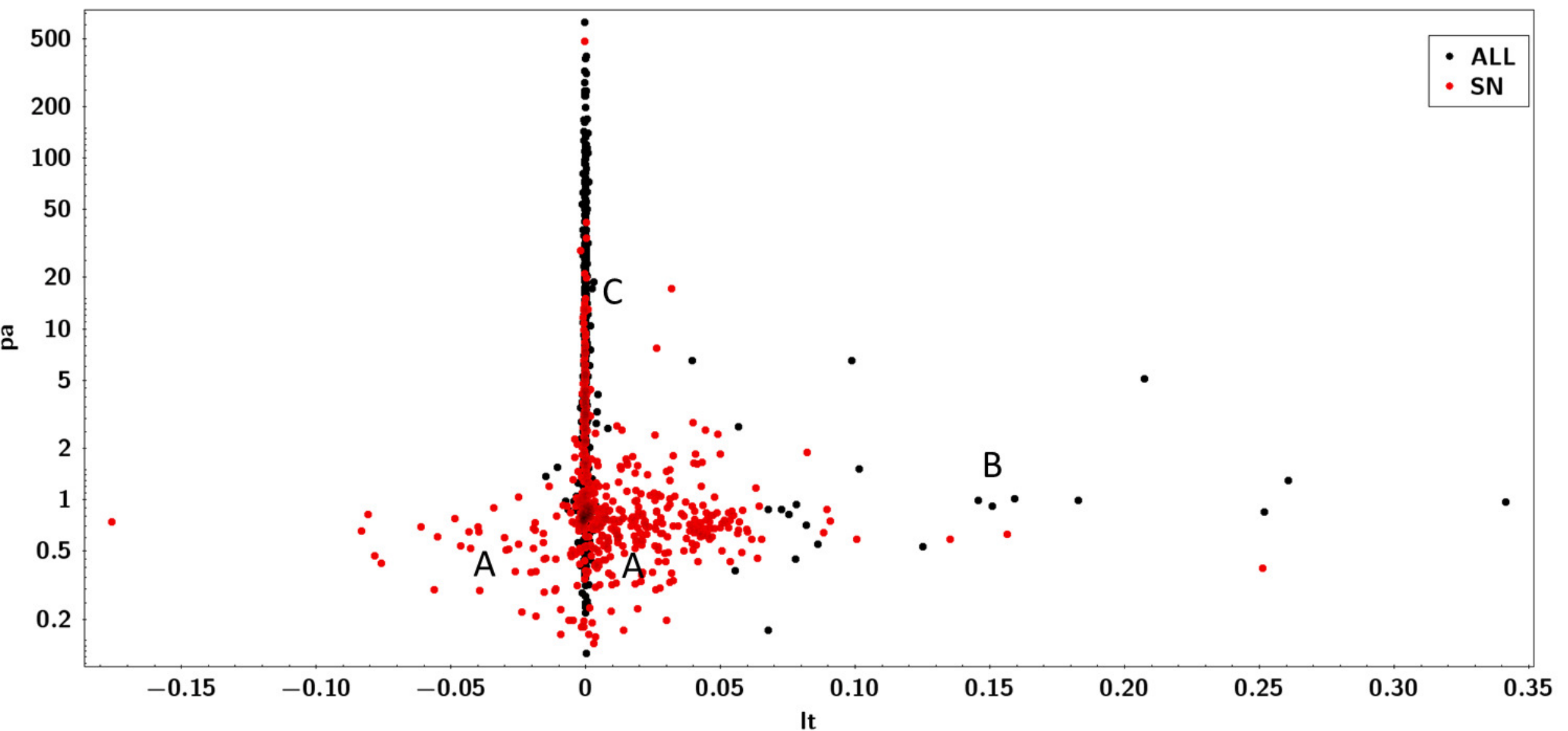}
        }\\
\subfigure[]{
            \includegraphics[width=0.80\textwidth]{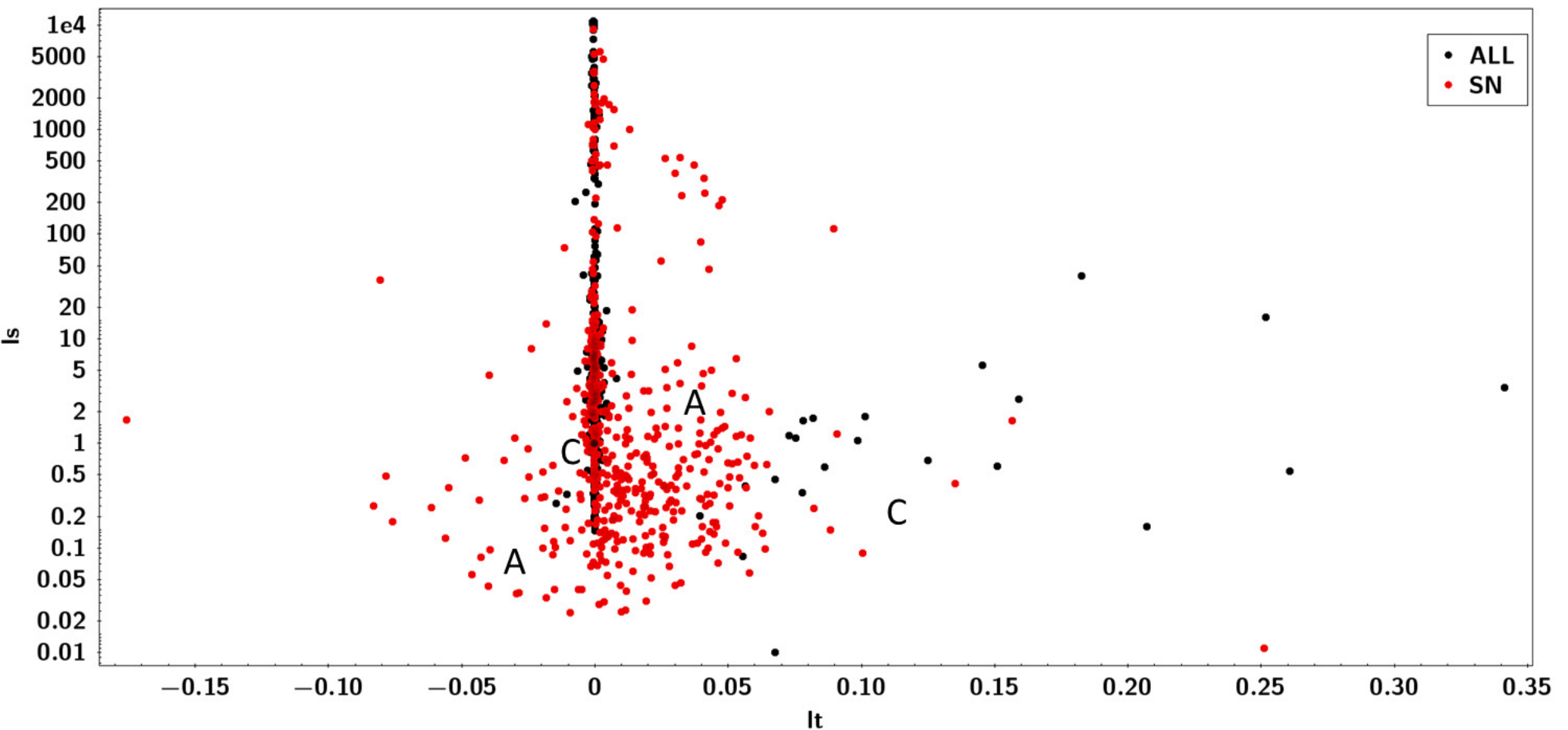}
        }\\
\end{center}
\caption{Panel (a): Comparison of features \textit{std} vs \textit{lt} in the case of \textit{SN vs ALL} experiment;
panel (b): the same plot but between \textit{pa} and \textit{lt} features; panel (c): the same plot but between \textit{ls} and \textit{lt} features.
Red color is related to \textit{SN} objects and black to \textit{ALL} class objects. The labels indicate, respectively, (A) pure \textit{SN} region
(i.e. a region populated only by \textit{SN} objects), (B) sparse \textit{SN} region (weak percentage of \textit{SN} objects), (C) mixed zone and (D) almost pure SN region. The overabundance of points having \textit{lt}$=0$
reflects the fact that RRL and AGN as well as any other impulsive variable have in average a constant behavior.}
\label{SN-lt-vs-std-pa}
\end{figure*}

\begin{figure*}
\begin{center}
\subfigure[]{
            \includegraphics[width=0.80\textwidth]{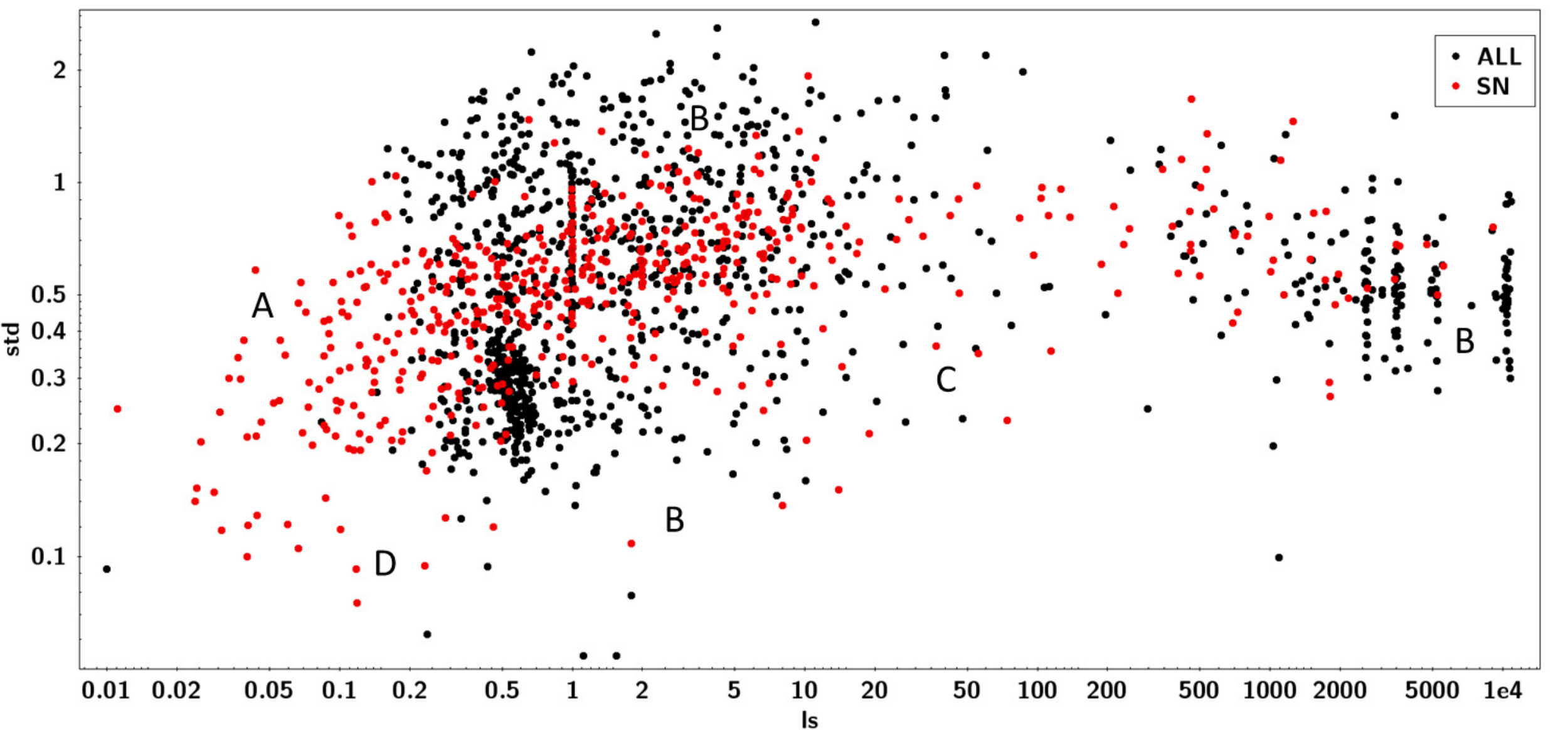}
        }\\
\subfigure[]{
            \includegraphics[width=0.80\textwidth]{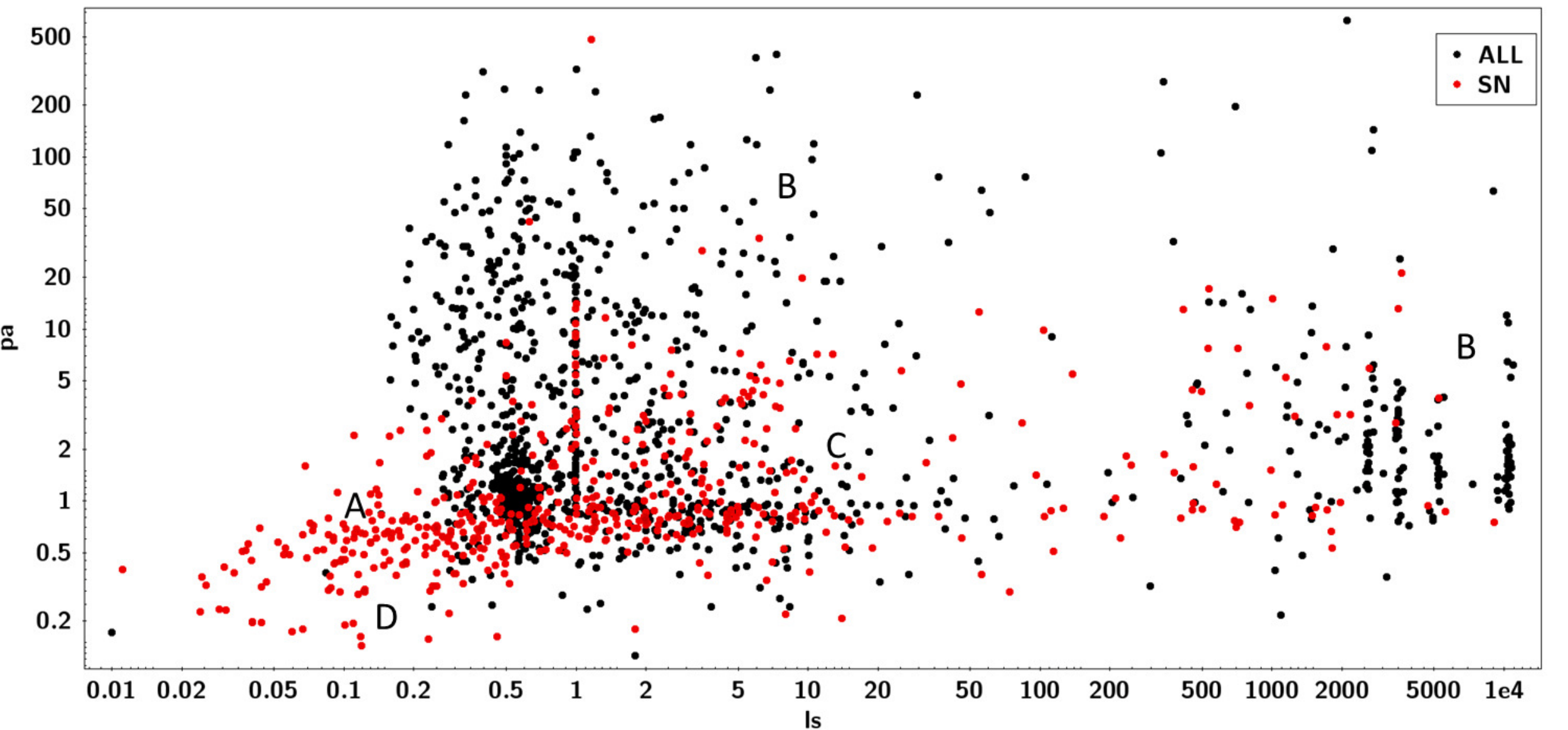}
        }\\
\subfigure[]{
            \includegraphics[width=0.80\textwidth]{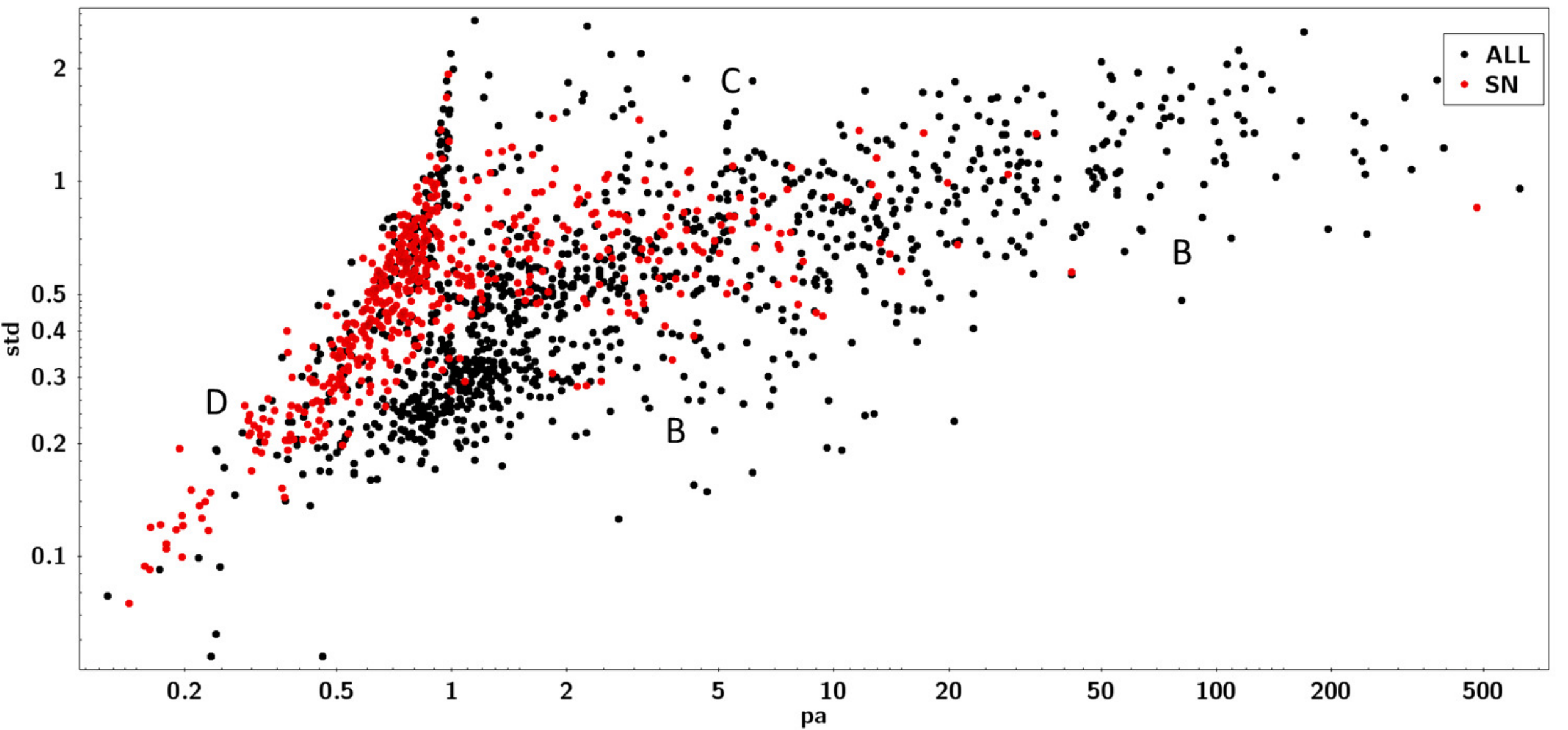}
        }\\
\end{center}
\caption{Panel (a): Comparison of features \textit{std} vs \textit{ls} in the case of \textit{SN vs ALL} experiment; panel (b): the same plot but between \textit{pa} and \textit{ls}
features; panel (c): the same plot but between \textit{std} and \textit{pa} features. Red color is related to \textit{SN} objects and black to \textit{ALL} class objects.
The labels indicate, respectively, (A) pure \textit{SN} region (i.e. a region populated only by \textit{SN} objects), (B) sparse \textit{SN} region (weak percentage of \textit{SN} objects),
(C) mixed zone and (D) almost pure SN region. The vertical structure at \textit{ls}$=1$ is an effect introduced by the sampling frequency of the survey (the structure is mainly populated by AGN, Bl and SN).}
\label{SN-ls-vs}
\end{figure*}

The capability to disentangle \textit{SN} class objects through the most relevant selected features appears evident by comparing them among each other.
In particular from figures \ref{SN-lt-vs-std-pa} and \ref{SN-ls-vs} it is possible to locate sub-regions entirely populated by \textit{SN} type objects (those labeled as A in the plots), as well as regions characterized by a weak (labeled as B) or strong (labeled as D) density of SN type objects. This implies that, besides the particular choice of the classifier, in the parameter space defined by the most relevant features there are combined ranges of feature distributions (for instance \textit{ls, pa, lt} and \textit{std}) able to classify \textit{SN} type objects from the rest of the data types with a high confidence. This evidence is also confirmed by the purity percentages obtained in the case of \textit{SN vs ALL} experiment by the three classifiers used.

%\newpage
\section{Conclusions}

This work focused on the use of three well tested machine learning methods, respectively RF (Random Forest), MLPQNA (Multi Layer Perceptron trained by the Quasi Newton learning rule) and KNN (K-Nearest Neighbors), to classify transient objects and
it is a first step towards a framework where different classifiers shall work in collaborative way on the same data to obtain a reliable, accurate and reproducible classification of variable objects.

We run a multi-class (all six object categories available) and derived three types of binary classification experiments:
(\textit{i}) \textit{Cataclysmic Variables} vs \textit{ALL} (\textit{AGN, SN, Fl, Bl} types);
(\textit{ii}) \textit{Extra-Galactic} (\textit{AGN} and \textit{Bl} types) vs \textit{Galactic} (\textit{CV}, \textit{SN} and \textit{Fl} types);
(\textit{iii}) \textit{SN vs ALL} (\textit{AGN, Bl, CV, Fl} and \textit{RRL} types).

Taking into account the results of the binary classification experiments only, the performance can be summarized as it follows: for the \textit{SN vs ALL} the best method is RF, which achieves a $\sim87\%$ efficiency, with a completeness of $\sim73\%$ and a purity for SNs of $\sim86\%$. In the same experiment the MLPQNA obtains a slightly higher purity ($\sim90\%$) at a price of a lower completeness ($\sim61\%$). In the \textit{Cataclysmic Variables} vs \textit{ALL} the best performance is achieved by the MLPQNA ($\sim86\%$ efficiency with a completeness of $\sim79\%$ and a purity for CVs of $\sim80\%$). It is however worth noticing that the combination of the outcome of the three models allows to achieve better performance ($\sim92\%$ efficiency for both experiments). Finally, in the third experiment (\textit{X-GAL vs GAL}) the best results were achieved by the RF model, obtaining $\sim92\%$ efficiency, with a \textit{X-GAL} class completeness of $\sim69\%$ and a purity of $\sim88\%$.

By exploiting the feature importance score provided by the RF model, the ranking between feature grouping and classification performance was investigated and it led to the identification of a special group of features which carry most information, regardless the specific experiment. This is a crucial issue since, in the big data regime which is typical of future surveys the identification of an optimal set of feature is needed in order to reduce computing time.

Overall, RF and MLPQNA achieve better results when the classifiers are used in combination.
The combined and hierarchical use of a wide set of classifiers could be finalized into a framework having as main purpose the capability to disentangle and identify the largest variety of variable objects \citep{donalek2}.

\section*{Acknowledgments}
The authors wish to thank the anonymous referee for the very helpful comments and suggestions which contributed to optimize the manuscript.
The authors wish to acknowledge financial support from the Italian Ministry of University and Research through
the grant PRIN-MIUR "Cosmology with Euclid" and from the Keck Institute for Space Studies who sponsored the working group on
TDA. MB and SC acknowledge financial contribution from the agreement ASI/INAF I/023/12/1.
Authors also wish to thank K. Polsterer for useful discussions.
This work made use of the CRTS public archive data, of the CTSCS service and of the DAMEWARE infrastructure. CD and SGD acknowledge a partial support from the NSF grants AST-1413600 and AST-1313422.  We are thankful to A. J. Drake, A. A. Mahabal, and M. J. Graham for their key contributions in the CRTS project.

%\newpage

\appendix
\section{Experiment tables}
\label{AppI}

\FloatBarrier
\begin{table}\scriptsize
\centering
\begin{tabular}{| c | c | c | c |}
\hline
\textbf{Statistics}&\textbf{All features}&\textbf{5 features}&\textbf{3 features}\\
\hline
\hline
\textit{Eff}&72.46&73.85&73.54\\
\hline
\textit{Comp CV}&71.43&73.63&72.53\\
\hline
\textit{Comp SN}&63.21&78.30&80.19\\
\hline
\textit{Comp Bl}&31.58&26.31&26.82\\
\hline
\textit{Comp AGN}&61.29&58.06&74.19\\
\hline
\textit{Comp Fl}&47.37&52.63&57.89\\
\hline
\textit{Comp RRL}&84.74&96.61&91.52\\
\hline
\textit{Pur CV}&57.52&71.28&74.16\\
\hline
\textit{Pur SN}&65.69&76.85&76.58\\
\hline
\textit{Pur Bl}&33.33&29.41&23.43\\
\hline
\textit{Pur AGN}&76.00&64.28&58.97\\
\hline
\textit{Pur Fl}&90.00&83.33&68.75\\
\hline
\textit{Pur RRL}&87.72&86.36&77.14\\
\hline
\end{tabular}
\caption{Results of the experiments with the MLPQNA for the \textit{six-class} experiment, obtained using the features in order of importance, following the list of Fig.~\ref{list_six}. All the results are in percentage.}
\label{six_MLPQNA}
\end{table}

\FloatBarrier
\begin{table}\scriptsize
\centering
\begin{tabular}{| c | c | c | c |}
\hline
\textbf{Statistics}&\textbf{All features}&\textbf{5 features}&\textbf{3 features}\\
\hline
\hline
\textit{Eff}&79.14&77.30&72.08\\
\hline
\textit{Comp CV}&79.12&79.12&68.13\\
\hline
\textit{Comp SN}&83.96&83.96&78.30\\
\hline
\textit{Comp Bl}&36.84&31.58&26.31\\
\hline
\textit{Comp AGN}&77.42&67.74&64.52\\
\hline
\textit{Comp Fl}&52.63&47.37&52.63\\
\hline
\textit{Comp RRL}&94.91&93.22&93.22\\
\hline
\textit{Pur CV}&74.22&73.47&66.67\\
\hline
\textit{Pur SN}&76.72&76.72&74.11\\
\hline
\textit{Pur Bl}&50.00&37.50&31.25\\
\hline
\textit{Pur AGN}&85.71&80.77&74.07\\
\hline
\textit{Pur Fl}&100.00&81.82&71.43\\
\hline
\textit{Pur RRL}&93.33&94.83&87.30\\
\hline
\end{tabular}
\caption{Results of the experiments with the Random Forest for the \textit{six-class} experiment, obtained using the features in order of importance, following the list of Fig.~\ref{list_six}. All the results are in percentage.}
\label{six_RF}
\end{table}

\FloatBarrier
\begin{table}\scriptsize
\centering
\begin{tabular}{| c | c | c | c |}
\hline
\textbf{Statistics}&\textbf{All features}&\textbf{5 features}&\textbf{3 features}\\
\hline
\hline
\textit{Eff}&55.38&66.77&61.54\\
\hline
\textit{Comp CV}&64.83&68.13&68.13\\
\hline
\textit{Comp SN}&62.26&72.64&58.49\\
\hline
\textit{Comp Bl}&15.79&10.53&5.26\\
\hline
\textit{Comp AGN}&61.29&61.29&48.39\\
\hline
\textit{Comp Fl}&10.53&36.84&47.37\\
\hline
\textit{Comp RRL}&52.54&84.74&76.27\\
\hline
\textit{Pur CV}&55.14&63.26&59.61\\
\hline
\textit{Pur SN}&55.46&66.38&65.38\\
\hline
\textit{Pur Bl}&16.67&10.00&12.50\\
\hline
\textit{Pur AGN}&70.37&70.37&57.69\\
\hline
\textit{Pur Fl}&25.00&87.50&52.94\\
\hline
\textit{Pur RRL}&67.39&89.28&68.18\\
\hline
\end{tabular}
\caption{Results of the experiments with the KNN for the \textit{six-class} experiment, obtained using the features in order of importance, following the list of Fig.~\ref{list_six}. All the results are in percentage.}
\label{six_KNN}
\end{table}

\FloatBarrier
\begin{table}\scriptsize
\centering
\begin{tabular}{| c | c | c | c | c | c | c |}
\hline
\textbf{Features}&\textbf{Eff}&\textbf{Comp1}&\textbf{Comp2}&\textbf{Pur1}&\textbf{Pur2}&\textbf{MCC}\\
\hline
\hline
\textit{20}&77.82&69.23&82.28&67.02&83.72&0.51\\
\hline
\textit{3}&79.70&54.94&92.57&79.36&79.80&0.53\\
\hline
\textit{{\bf5}}&{\bf82.71}&{\bf70.33}&{\bf89.14}&{\bf77.11}&{\bf85.24}&{\bf0.61}\\
\hline
\textit{6}&79.70&67.03&86.28&71.76&83.42&0.54\\
\hline
\textit{9}&80.07&73.63&83.43&69.79&85.88&0.56\\
\hline
\textit{10}&77.82&72.53&80.57&66.00&84.94&0.52\\
\hline
\textit{11}&79.70&73.63&82.86&69.07&85.80&0.56\\
\hline
\textit{{\bf5*}}&{\bf86.09}&{\bf79.12}&{\bf89.71}&{\bf80.00}&{\bf89.20}&{\bf0.69}\\
\hline
\end{tabular}
\caption{Results of the experiments with the MLPQNA for the \textit{CV (class 1) vs ALL (class 2)} classification, obtained using the features in order of importance, following the list of Fig.~\ref{list_cv}. All the results are in percentage, except the MCC. The last row (5*) refers to the best result, obtained with an optimization of the model configuration parameters.}
\label{cv_mlp}
\end{table}

\begin{table}\scriptsize
\centering
\begin{tabular}{| c | c | c | c | c | c | c |}
\hline
\textbf{Features}&\textbf{Eff}&\textbf{Comp1}&\textbf{Comp2}&\textbf{Pur1}&\textbf{Pur2}&\textbf{MCC}\\
\hline
\hline
\textit{{\bf20}}&{\bf84.02}&{\bf70.01}&{\bf91.81}&{\bf81.96}&{\bf85.12}&{\bf0.64}\\
\hline
\textit{3}&77.47&60.49&86.87&71.18&80.46&0.49\\
\hline
\textit{5}&83.04&71.57&89.38&78.17&85.50&0.62\\
\hline
\textit{6}&83.49&71.83&89.89&79.21&85.62&0.63\\
\hline
\textit{9}&84.85&74.46&90.74&81.09&86.84&0.66\\
\hline
\textit{10}&85.08&74.73&90.86&81.28&86.99&0.67\\
\hline
\textit{11}&84.32&72.97&90.63&80.55&86.21&0.65\\
\hline
\end{tabular}
\caption{Results of the experiments with the RF for the \textit{CV (class $1$) vs ALL (class $2$)} classification, obtained using the features in order of importance, following the list of Fig.~\ref{list_cv} and a cross validation with $k = 10$. All the results are expressed as  percentages, except the MCC.}
\label{cv_rf}
\end{table}

\begin{table}\scriptsize
\centering
\begin{tabular}{| c | c | c | c | c | c | c |}
\hline
\textbf{Features}&\textbf{Eff}&\textbf{Comp1}&\textbf{Comp2}&\textbf{Pur1}&\textbf{Pur2}&\textbf{MCC}\\
\hline
\hline
\textit{20}&78.07&57.75&89.04&73.71&79.82&0.50\\
\hline
\textit{3}&77.32&59.80&86.94&71.42&80.19&0.49\\
\hline
\textit{5}&78.07&67.95&83.62&68.91&82.98&0.52\\
\hline
\textit{{\bf6}}&{\bf79.65}&{\bf70.04}&{\bf84.96}&{\bf71.55}&{\bf84.00}&{\bf0.55}\\
\hline
\textit{9}&78.82&64.75&86.51&71.90&82.08&0.52\\
\hline
\textit{10}&78.75&63.55&87.03&72.33&81.79&0.52\\
\hline
\textit{11}&78.22&61.87&87.23&72.14&81.10&0.51\\
\hline
\end{tabular}
\caption{Results of the experiments with the KNN for the \textit{CV (class $1$) vs ALL (class $2$)} classification, obtained using the features in order of importance, following the list of Fig.~\ref{list_cv} and a cross validation with $k = 10$. All the results are expressed as percentages, except the MCC.}
\label{cv_knn}
\end{table}

\begin{table}\scriptsize
\centering
\begin{tabular}{| c | c | c | c | c | c | c |}
\hline
\textbf{Features}&\textbf{Eff}&\textbf{Comp1}&\textbf{Comp2}&\textbf{Pur1}&\textbf{Pur2}&\textbf{MCC}\\
\hline
\hline
\textit{20}&87.97&66.00&93.05&68.75&92.20&0.60\\
\hline
\textit{{\bf5}}&{\bf88.34}&{\bf72.00}&{\bf92.13}&{\bf67.92}&{\bf93.43}&{\bf0.63}\\
\hline
\textit{10}&86.09&72.00&89.35&61.02&93.24&0.58\\
\hline
\textit{5\dag}&88.34&66.00&93.52&70.21&92.24&0.61\\
\hline
\textit{{\bf5*}}&{\bf88.72}&{\bf68.00}&{\bf93.52}&{\bf70.83}&{\bf92.66}&{\bf0.62}\\
\hline
\textit{5\dag *}&88.72&66.00&93.98&71.74&92.27&0.62\\
\hline
\end{tabular}
\caption{Results of the experiments with the MLPQNA for the \textit{X-GAL (class $1$) vs GAL (class $2$)} classification, obtained using the features in order of importance, following the list of Fig.~\ref{list_xgal}. All the results are in percentage except the MCC. The row (5\dag) is referred to the features selected in the \textit{CV vs ALL} experiment. The 5* is the best result obtained by optimizing the model parameters, while the last row (5\dag *) is the best result obtained in the case of \textit{CV vs ALL} experiments.}
\label{evg_mlp}
\end{table}

\begin{table}\scriptsize
\centering
\begin{tabular}{| c | c | c | c | c | c | c |}
\hline
\textbf{Features}&\textbf{Eff}&\textbf{Comp1}&\textbf{Comp2}&\textbf{Pur1}&\textbf{Pur2}&\textbf{MCC}\\
\hline
\hline
\textit{20}&91.41&66.69&97.64&87.87&92.17&0.71\\
\hline
\textit{5}&90.73&66.40&96.90&84.50&92.04&0.69\\
\hline
\textit{{\bf10}}&{\bf91.71}&{\bf68.51}&{\bf97.55}&{\bf88.19}&{\bf92.58}&{\bf0.73}\\
\hline
\textit{5\dag}&88.47&59.91&95.46&76.37&90.64&0.61\\
\hline
\end{tabular}
\caption{Results of the experiments with the RF for the \textit{X-GAL (class $1$) vs GAL (class $2$)} classification, obtained using the features in order of importance, following the list of Fig.~\ref{list_xgal}, and a cross validation with $k = 10$. All the results are in percentage except the MCC. The last row (5\dag) is referred to the features selected in the \textit{CV vs ALL} experiment.}
\label{evg_rf}
\end{table}

\begin{table}\scriptsize
\centering
\begin{tabular}{| c | c | c | c | c | c | c |}
\hline
\textbf{Features}&\textbf{Eff}&\textbf{Comp1}&\textbf{Comp2}&\textbf{Pur1}&\textbf{Pur2}&\textbf{MCC}\\
\hline
\hline
\textit{20}&89.83&71.83&94.44&76.30&92.98&0.68\\
\hline
\textit{5}&87.80&64.68&93.73&72.19&91.35&0.61\\
\hline
\textit{{\bf10}}&{\bf90.05}&{\bf70.33}&{\bf95.32}&{\bf78.71}&{\bf92.59}&{\bf0.68}\\
\hline
\textit{5\dag}&86.66&65.09&91.96&67.46&91.39&0.58\\
\hline
\end{tabular}
\caption{Results of the experiments with the KNN for the \textit{X-GAL (class $1$) vs GAL (class $2$)} classification, obtained using the features in order of importance, following the list of Fig.~\ref{list_xgal}, and a cross validation with $k = 10$. All the results are in percentage except the MCC. The last row (5\dag) is referred to the features selected in the \textit{CV vs ALL} experiment.}
\label{evg_knn}
\end{table}

\begin{table}\scriptsize
\centering
\begin{tabular}{| c | c | c | c | c | c | c |}
\hline
\textbf{Features}&\textbf{Eff}&\textbf{Comp1}&\textbf{Comp2}&\textbf{Pur1}&\textbf{Pur2}&\textbf{MCC}\\
\hline
\hline
\textit{20}&80.00&68.87&85.39&69.52&85.00&0.54\\
\hline
\textit{5\dag}&85.23&71.70&91.78&80.85&87.01&0.66\\
\hline
\textit{{\bf3}}&{\bf84.92}&{\bf62.26}&{\bf95.89}&{\bf88.00}&{\bf84.00}&{\bf0.65}\\
\hline
\textit{5}&85.23&72.64&91.32&80.21&87.34&0.66\\
\hline
\textit{10}&82.15&77.36&84.47&70.69&88.52&0.60\\
\hline
\textit{{\bf3*}}&{\bf85.23}&{\bf61.32}&{\bf96.80}&{\bf90.28}&{\bf83.79}&{\bf0.66}\\
\hline
\end{tabular}
\caption{Results of the experiments with the MLPQNA for the \textit{SN (class $1$) vs ALL (class $2$)} classification, obtained using the features in order of importance, following the list of Fig.~\ref{list_sn}. All the results are in percentage except the MCC. The last column (3*) is referred to the best results obtained by an optimization of the model parameters.}
\label{sn_mlp}
\end{table}

\begin{table}\scriptsize
\centering
\begin{tabular}{| c | c | c | c | c | c | c | c | c |}
\hline
\textbf{Features}&\textbf{Eff}&\textbf{Comp1}&\textbf{Comp2}&\textbf{Pur1}&\textbf{Pur2}&\textbf{MCC}\\
\hline
\hline
\textit{20}&86.60&71.84&93.62&84.47&87.26&0.68\\
\hline
\textit{5\dag}&85.98&72.23&92.76&82.67&87.14&0.67\\
\hline
\textit{3}&85.30&69.74&92.77&82.27&86.26&0.65\\
\hline
\textit{5}&86.54&72.53&93.27&83.71&87.46&0.68\\
\hline
\textit{{\bf10}}&{\bf87.34}&{\bf72.81}&{\bf94.25}&{\bf86.00}&{\bf87.72}&{\bf0.70}\\
\hline
\end{tabular}
\caption{Results of the experiments with the RF for the \textit{SN (class $1$) vs ALL (class $2$)} classification, obtained using the features in order of importance, following the list of Fig.~\ref{list_sn}, and a cross validation with $k = 10$. All the results are in percentage except the MCC.}
\label{sn_rf}
\end{table}

\begin{table}\scriptsize
\centering
\begin{tabular}{| c | c | c | c | c | c | c |}
\hline
\textbf{Features}&\textbf{Eff}&\textbf{Comp1}&\textbf{Comp2}&\textbf{Pur1}&\textbf{Pur2}&\textbf{MCC}\\
\hline
\hline
\textit{20}&76.47&57.98&85.92&66.89&80.33&0.45\\
\hline
\textit{5\dag}&82.03&63.37&91.21&78.38&83.40&0.58\\
\hline
\textit{{\bf3}}&{\bf83.32}&{\bf66.33}&{\bf91.58}&{\bf79.38}&{\bf84.66}&{\bf0.61}\\
\hline
\textit{5}&79.87&59.39&89.89&74.00&81.81&0.52\\
\hline
\textit{10}&79.25&65.67&85.81&69.25&83.58&0.52\\
\hline
\end{tabular}
\caption{Results in percentage, except the MCC, of the experiments with the different groups of features from Fig.~\ref{list_sn}, obtained using the KNN for the classification \textit{SN (class $1$) vs ALL (class $2$)} and a cross validation with $k = 10$.}
\label{sn_knn}
\end{table}
\FloatBarrier

\label{lastpage}

\begin{thebibliography}{100}
% \bibitem[\protect\citeauthoryear{Baade and Zwicky}{1934}]{baade} Baade W., Zwicky F., Supernovae and Cosmic rays Physical
%Review 45, 138, 1934
% \bibitem[\protect\citeauthoryear{Bernardini}{2011}]{bernardini} Bernardini E., Astronomy in the Time Domain, Vol. 331,
%Science, February 2011
\bibitem[\protect\citeauthoryear{Bloom and Richards}{2011}]{bloom} Bloom J.S., Richards J.W., 2011, Data Mining and Machine-Learning in Time-Domain Discovery \& Classification, "Advances in Machine Learning and Data Mining for Astronomy", arXiv:1104.3142
\bibitem[\protect\citeauthoryear{Breiman}{2001}]{breiman2001} Breiman, L., Random Forests. Machine Learning, Springer Eds., 45, 1, 25-32 (2001)
% \bibitem[\protect\citeauthoryear{Brescia et al.}{2009}]{brescia2009} Brescia, M., et al., DAME: A Distributed Web Based
%Framework for Knowledge Discovery in Databases, Mem. SAIt Suppl, 13, 56, 2009
% \bibitem[\protect\citeauthoryear{Brescia et al.}{2011}]{brescia2011} Brescia M., Cavuoti S., D'Abrusco R., Laurino O., Longo
%G., V International Workshop on Distributed Cooperative Laboratories:
%Instrumenting the Grid, in Remote Instrumentation for eScience and Related
%Aspects, F. Davoli et al. (eds.), Springer:NY, 2011
\bibitem[\protect\citeauthoryear{Brescia et al.}{2012}]{brescia2} Brescia et al., The detection of globular clusters in
galaxies as a data mining problem, Monthly Notices of the Royal Astronomical
Society, 421, 2, 1155-1165, 2012
 \bibitem[\protect\citeauthoryear{Brescia et al.}{2014}]{dame} Brescia M., Cavuoti S., Longo G., et al., 2014, PASP, 126, 942, 783-797
% \bibitem[\protect\citeauthoryear{Butler et al.}{2011}]{butler} Butler N.R., Bloom J.S., Optimal time-series selection of
%Quasars, AJ, 141, 93, 2011
 \bibitem[\protect\citeauthoryear{Byrd et al.}{1994}]{byrd1994} Byrd, R.H, Nocedal, J., and Schnabel, R.B., Mathematical Programming, 63, 129 (1994)

 \bibitem[\protect\citeauthoryear{Castillo et al.}{1997}]{castillo}Castillo, E., Gutierrez, J. M. and Hadi, A. S. (1997). "Learning Bayesian Networks". Expert Systems and Probabilistic Network Models. Monographs in computer science. New York: Springer-Verlag. pp. 481-528. ISBN 0-387-94858-9.

% \bibitem[\protect\citeauthoryear{Cavuoti et al.}{2012}]{cavuoti} Cavuoti S., Brescia M., Longo G., Garofalo M., Nocella A.,
%2012, DAME: A Web Oriented Infrastructure for Scientific Data Mining and
%Exploration, Science - World Scientific Publishing Co. Pte. Ltd., ISBN
%9789814383295, pp. 241-247, 2012
\bibitem[\protect\citeauthoryear{Cavuoti et al.}{2014}]{cavuoti2014} Cavuoti S.; Brescia M.; D'Abrusco R.; Longo G. \& Paolillo M., 2014, MNRAS 437, 968
\bibitem[\protect\citeauthoryear{Cavuoti et al.}{2015}]{cavuoti2015} Cavuoti S., Brescia M., De Stefano, V., Longo G., 2015, Experimental Astronomy, Springer, in press, eprint arXiv:1501.06506
\bibitem[\protect\citeauthoryear{Chang \& Lin}{2011}]{chang2011}  Chang, Chih-Chung and  Lin, Chih-Jen, 2011, LIBSVM: A library for support vector machines, ACM Transactions on Intelligent Systems and Technology, 2, 27.

\bibitem[\protect\citeauthoryear{Closson Ferguson}{2015}]{closson2015} Closson Ferguson, H., 2015, IAU General Assembly, Meeting 29, 2257590
 %\bibitem[\protect\citeauthoryear{Chow et al.}{2005}]{chow} Chow-Choong Ngeow, Shashi M. Kanbur, The linearity of the Wesenheit function for the Large Magellanic Cloud Cepheids, Monthly Notice of  the Royal Astronomical Society, 360, S. 1033-1039, 2005
 %\bibitem[\protect\citeauthoryear{Cox}{1980}]{cox} Cox J.P., Theory of stellar pulsation, Princeton, N.J.: Princeton Univ. Press., 1980
 \bibitem[\protect\citeauthoryear{Debosscher et al.}{2007}]{debosscher} Debosscher J. et al., Automated supervised classification
of variable stars, A\&A, 475, 1159, 2007
% \bibitem[\protect\citeauthoryear{Djorgovski et al.}{2008}]{djorgovski} Djorgovski et al., The Palomar-Quest Digital Synoptic Sky
%Survey, Astron. Nach., 329, 263, 2008
 \bibitem[\protect\citeauthoryear{Donalek et al.}{2013}]{donalek2} Donalek C., Djorgovski G., Longo G. et al., Machine
Learning Techniques in the Time Domain, Naples, May 2013
\bibitem[\protect\citeauthoryear{Drake et al.}{2010}]{drake2010} Drake, A.J. et al., 2010, ApjL, 718, 127
\bibitem[\protect\citeauthoryear{Drake et al.}{2009}]{drake} Drake A.J. et al., First Results from the Catalina Real-time
Transient Survey, ApJ, 696, 870, 2009
% \bibitem[\protect\citeauthoryear{Dubath et al.}{2011}]{dubath2} Dubath P., Rimoldini L., Suveges M. et al., Random forest
%automated supervised classification of Hipparcos periodic variable stars, Mon.
%Not. R. Astron. Soc., May, 651, 2011
 \bibitem[\protect\citeauthoryear{Dubath}{2012}]{dubath1} Dubath P., Hipparcos Variable Star Detection and
Classification Efficiency, Astrostatistics and Data Mining, Springer Series in
Astrostatistics, Volume 2. ISBN 978-1-4614-3322-4. Springer Science+Business
Media New York, p. 117, 2012
\bibitem[\protect\citeauthoryear{du Buisson et al.}{2014}]{dubuisson2014} du Buisson, L., et al., 2014, arXiv:1407.4118
% \bibitem[\protect\citeauthoryear{Eddington}{1918}]{eddington} Eddington A.S., Stars, Gaseous, On the pulsations of a gaseous star, MNRAS, Vol. 79, p.2-22, 1918
 \bibitem[\protect\citeauthoryear{Eyer and Mowlavi}{2007}]{eyer} Eyer L., Mowlavi N., 2008, Variable Stars across the HR
Diagram, JPhCS 118
\bibitem[\protect\citeauthoryear{Geisser}{1975}]{geisser1975} Geisser, S., 1975. The predictive sample reuse method with applications. Journal of the American Statistical Association, 70 (350), 320-328.
\bibitem[\protect\citeauthoryear{Goldstein et al.}{2015}]{goldstein2015} Goldstein, D.A. et al., 2015, AJ, 150, 3, 82, 15 pp.
\bibitem[\protect\citeauthoryear{Graham et al.}{2015}]{graham2015} Graham M.J. et al., 2015, MNRAS, 453, 2, 1562-1576
\bibitem[\protect\citeauthoryear{Graham et al.}{2012}]{graham2012} Graham M.J. et al., 2012a, Proc. of SPIE, 8448, 84480P, 8 pages
\bibitem[\protect\citeauthoryear{Graham et al.}{2012}]{graham} Graham M.J. et al., 2012b, Data challenges of time domain astronomy, To appear on special issue of Distributed and Parallel Databases on Data Intensive eScience, arXiv:1208.2480
% \bibitem[\protect\citeauthoryear{Grindlay et al.}{2012}]{grindlay} Grindlay J. et al., Opening the 100-year window for
%time-domain astronomy - New Horizons in Time-Domain Astronomy, Proceedings of
%the International Astronomical Union, IAU Symposium, Volume 285, p. 29-34
% \bibitem[\protect\citeauthoryear{Harwit}{2003}]{harwit} Harwit M., The Growth of Astrophysical Understanding, Phys.
%Today, 56, 38, 2003
\bibitem[\protect\citeauthoryear{Hanley and McNeil}{1982}]{hanley} Hanley, J.~A., McNeil, B.~J., 1982, The Meaning and Use of the Area under a Receiver Operating Characteristic (ROC) Curve. Radiology, 143, (1), 29–36
\bibitem[\protect\citeauthoryear{Hastie et al.}{2001}]{hastie} Hastie T., Tibshirani R., Friedman J.H., 2001, The Elements of Statistical Learning: Data Mining, Inference, and Prediction, Springer
% \bibitem[\protect\citeauthoryear{Hertzprung}{1909}]{hr1} Hertzprung E., Uber die Sterne der Unterabteilung c und ac nach der Spektralklassifikation von Antonia C. Maury, Astronomische Nachrichten 179 (4296): 373-380, 1909
% \bibitem[\protect\citeauthoryear{Hey et al.}{2009}]{paradigm} Hey T., Tansley S., Tolle K., The Fourth Paradigm, Microsoft Research
\bibitem[\protect\citeauthoryear{Kitty et al.}{2014}]{kitty}  Kitty K. Lo, Tara Murphy, Umaa Rebbapragada, Kiri Wagstaff, 2013, Online Classification for Time-Domain Astronomy, ICDM Workshops 2013: 24-31
\bibitem[\protect\citeauthoryear{Kohonen}{2007}]{kohonen2007}Kohonen, T.; 2007, Self-Organizing Maps, Springer, Heidelberg, Second ed.; Vol. 30.

%\bibitem[\protect\citeauthoryear{Law et al.}{2009}]{law1} Law et al., The Palomar Transient Factory: System Overview,
%Performance and First Results, PASP 121 1395L
%\bibitem[\protect\citeauthoryear{Law et al.}{2010}]{law2} Law et al., The Palomar Transient Factory Survey Camera: 1st
%Year Performance and Results, SPIE 7735, 2010
 %\bibitem[\protect\citeauthoryear{Longo and Brescia}{2012}]{longo} Longo G., Brescia M., Time Domain Astronomy: a new frontier of astronomy, Interdepartmental Physics-Mathematics lecture, Department of Physics, University Federico II, Napoli, June 7, 2012
\bibitem[\protect\citeauthoryear{Matthews}{1975}]{Matthews} Matthews, B. W., "Comparison of the predicted and observed secondary structure of T4 phage lysozyme". Biochimica et Biophysica Acta (BBA) - Protein Structure 405 (2): 442–451, 1975
\bibitem[\protect\citeauthoryear{McCulloch and Pitts}{1943}]{pitts} McCulloch W.S., Pitts W., Bullettin of Mathematical Biophysics 5:115-133, 1943
\bibitem[\protect\citeauthoryear{McLachlan and Peel}{2001}]{McLachlan} McLachlan, G., and Peel, D.. Finite Mixture Models. Hoboken, NJ: John Wiley \& Sons, Inc., 2000.
 %\bibitem[\protect\citeauthoryear{Minkowski}{1941}]{minkovski} Minkowski R., Spectra of Supernovae, Publications of the Astronomical Society of the Pacific, Vol. 53, No. 314, pp. 224-225, August 1941
% \bibitem[\protect\citeauthoryear{Minsky and Papert}{1969}]{minsky} Minsky M.L., Papert S.A., Perceptrons, Cambridge, MA: MIT
%Press, 1969
 \bibitem[\protect\citeauthoryear{Pedregosa et al.}{2011}]{pedregosa} Pedregosa et al., 2011, Scikit-learn: Machine Learning in Python, JMLR 12, pp. 2825-2830.
 %\bibitem[\protect\citeauthoryear{Phillips}{1993}]{phillips} Phillips M.M., The absolute magnitudes of Type IA supernovae, ApJ, Part 2 - Letters (ISSN 0004-637X), vol. 413, no. 2, p. L105-L108, 1993
\bibitem[\protect\citeauthoryear{Provost et al.}{1998}]{provost1998}Provost, F., Fawcett, T., Kohavi, R., 1998. The Case Against Accuracy Estimation for Comparing Induction Algorithms. Proceedings of the 15th International Conference on Machine Learning. Morgan Kaufmann. pp. 445-553.
 \bibitem[\protect\citeauthoryear{Richards et al.}{2011}]{richards} Richards J.W. et al., Variable star classification, ApJ,
733, 1; arXiv:1101.1959, 2011
\bibitem[\protect\citeauthoryear{Rebbapragada}{2014}]{rebbapragada2014} Rebbapragada, U., 2014, Data Triage of Astronomical Transients: A Machine Learning Approach. The Third Hot-wiring the Transient Universe Workshop, Ed. P.R. Wozniak et al., November 2013
 %\bibitem[\protect\citeauthoryear{Russell}{1914}]{hr2} Russell H.N., Relations Between the Spectra and Other Characteristics of the Stars, Popular Astronomy 22: 275-294, 1914
% \bibitem[\protect\citeauthoryear{Sako et al.}{2008}]{sako} Sako M. et al., The Sloan Digital Sky Survey-II Supernova
%Survey: Search Algorithm and Follow-up Observations, AJ, 135:348-373, January
%2008
 \bibitem[\protect\citeauthoryear{Scargle}{1982}]{scargle} Scargle J.D., Studies in astronomical time series analysis.
II - Statistical aspects of spectral analysis of unevenly spaced data, ApJ,
263, 835, 1982
% \bibitem[\protect\citeauthoryear{Schwarzenberg-Czerny}{1989}]{czerny} Schwarzenberg-Czerny A., On the advantage of using analysis
%of variance for period search, MNRAS, 241, 153, 1989
% \bibitem[\protect\citeauthoryear{Spearman}{1904}]{spearman} Spearman C., The proof and measurement of association
%between two things, Amer. J. Psychol. 15: 72–101, 1904
% \bibitem[\protect\citeauthoryear{Stetson}{1996}]{stetson} Stetson P.B., On the Automatic Determination of Light-Curve
%Parameters for Cepheid Variables, Publ. Astron. Soc. Pacific 108, 851, 1996
 \bibitem[\protect\citeauthoryear{Wright et al}{2015}]{wright} Wright, D. E., Smartt,  S. J., Smith K. W. et al. MNRAS 449 (1): 451-466, 2015

 \bibitem[\protect\citeauthoryear{Yahya et al.}{2015}]{yahya2015} Yahya, S. et al., 2015, MNRAS, 450, 3, 2251-2260

 \end{thebibliography}
\end{document}